\documentclass{aa}
\usepackage{psfig} 
\usepackage{amssymb}
\psfigurepath{fig}
\usepackage{graphics} 

\begin{document}
\thesaurus{03         
             (11.05.2;   
              13.09.1;   
              11.19.3;   
              11.19.1)   
}

\title{The bulk of the cosmic infrared background resolved by ISOCAM}
\author{\bf D.~Elbaz\inst{1,2,3}
\and C.J.~Cesarsky\inst{1,4}
\and P.~Chanial\inst{1}
\and H.~Aussel\inst{1,5}
\and A.~Franceschini\inst{6}
\and D.~Fadda\inst{1,7}
\and R.R.~Chary\inst{3}
}
\offprints{delbaz@cea.fr}
\institute{
DAPNIA/Service d'Astrophysique, CEA/Saclay, 91191 Gif-sur-Yvette Cedex, France \and
Physics Department, University of California, Santa Cruz, CA 95064, USA \and
Department of Astronomy \& Astrophysics, University of California, Santa Cruz, CA 95064, USA \and
European Southern Observatory, Karl-Schwarzchild-Strasse, 2 D-85748 Garching bei Muenchen, Germany \and
Institute For Astronomy, 2680 Woodlawn Drive, Honolulu, Hawaii 96822, USA \and
Dipartimento di Astronomia, Vicolo Osservatorio 2, I-35122 Padova, Italy \and
Instituto de Astrofisica de Canarias, Via Lactea, S/N E38200, La Laguna (Tenerife), SPAIN}

\date{Received: ; Accepted: }

\maketitle
\authorrunning{D. Elbaz et al}
\titlerunning{The bulk of the cosmic infrared background resolved by ISOCAM}
\markboth{D. Elbaz et al}{The bulk of the cosmic infrared background resolved by ISOCAM}

\begin{abstract}
Deep extragalactic surveys with ISOCAM revealed the presence of a
large density of faint mid-infrared (MIR) sources. We have computed
the 15\,$\mu$m integrated galaxy light produced by these galaxies
above a sensitivity limit of 50 $\mu$Jy. It sets a lower limit to the
15\,$\mu$m extragalactic background light of (2.4 $\pm$ 0.5) nW
m$^{-2}$ Hz$^{-1}$.

The redshift distribution of the ISOCAM galaxies is inferred from the
spectroscopically complete sample of galaxies in the Hubble Deep Field
North (HDFN). It peaks around $z\sim$ 0.8 in agreement with studies in
other fields. The rest-frame 15\,$\mu$m and bolometric infrared
(8-1000\,$\mu$m) luminosities of ISOCAM galaxies are computed using
the correlations that we establish between the 6.75, 12, 15\,$\mu$m
and infrared (IR) luminosities of local galaxies. The resulting IR
luminosities were double-checked using radio (1.4 GHz) flux densities
from the ultra-deep VLA and WSRT surveys of the HDFN on a sample of 24
galaxies as well as on a sample of 109 local galaxies in common
between ISOCAM and the NRAO VLA Sky Survey (NVSS). This comparison
shows for the first time that MIR and radio luminosities correlate up
to $z\sim$ 1. This result validates the bolometric IR luminosities
derived from MIR luminosities unless both the radio-far infrared (FIR)
and the MIR-FIR correlations become invalid around $z\sim$ 1.

The fraction of IR light produced by active nuclei was computed from
the cross-correlation with the deepest X-ray surveys from the Chandra
and XMM-Newton observatories in the HDFN and Lockman Hole
respectively. We find that at most 20\,$\%$ of the 15\,$\mu$m
integrated galaxy light is due to active galactic nuclei (AGNs) unless
a large population of AGNs was missed by Chandra and XMM-Newton.

About 75\,$\%$ of the ISOCAM galaxies are found to belong to the class
of luminous infrared galaxies ($L_{\rm IR}$ $\geq$ $10^{11}$
$L_{\sun}$). They exhibit star formation rates of the order of $\sim$
100 $M_{\sun}$ $yr^{-1}$. The comoving density of infrared light due
to these luminous IR galaxies was more than 40 times larger at $z\sim$
1 than today.

The contribution of ISOCAM galaxies to the peak of the cosmic infrared
background (CIRB) at 140\,$\mu$m was computed from the MIR-FIR
correlations for star forming galaxies and from the spectral energy
distribution of the Seyfert 2, NGC 1068, for AGNs. We find that the
galaxies unveiled by ISOCAM surveys are responsible for the bulk of
the CIRB, i.e (16 $\pm$ 5) nW m$^{-2}$ Hz$^{-1}$ as compared to the
(25 $\pm$ 7) nW m$^{-2}$ Hz$^{-1}$ measured with the COBE satellite,
with less than 10\,$\%$ due to AGNs. Since the CIRB contains most of
the light radiated over the history of star formation in the universe,
this means that a large fraction of present-day stars must have formed
during a dusty starburst event similar to those revealed by ISOCAM.
\end{abstract}
\section{Introduction}
The extragalactic background light (EBL) is a measurement of the sum
of the light produced by all extragalactic sources over cosmic
time. When it is integrated over the full spectral range, the
so-called cosmic background is a fossil record of the overall activity
of all galaxies from their birth until now. It can be considered as
the global energetic budget available for any model aiming at
simulating the birth and fate of galaxies during the Hubble
time. However the physical origin of this light will remain unknown
until we have pinpointed the individual sources responsible for
it. The goal of the present paper is to demonstrate that an important
new result has come from the combination of a series of deep
extragalactic surveys performed in the mid-infrared (MIR) at
15\,$\mu$m with the ISOCAM camera (Cesarsky et al. 1996a) onboard the
Infrared Space Observatory (ISO, Kessler et al. 1996): we suggest here
that the galaxies detected in these surveys, which median redshift of
$z\sim 0.8$ was measured from a sub-sample of ISOCAM galaxies,
contribute dominantly to the cosmic infrared background (CIRB),
i.e. the EBL integrated over all wavelengths within $\lambda$=
5 to 1000\,$\mu$m.

The CIRB was recently detected and measured thanks to the cosmic
background explorer (COBE) instruments FIRAS (Far Infrared Absolute
Spectrometer) and DIRBE (Diffuse Infrared Background Experiment)
(Puget et al. 1996, Fixsen et al. 1998, Lagache et al.  1999, 2000,
Hauser et al. 1998, Dwek et al. 1998, Finkbeiner et al. 2000) from
100\,$\mu$m to 1 mm. It peaks around $\lambda_{max}\simeq$ 140\,$\mu$m
and was found to represent at least half and maybe two thirds of the
overall cosmic background (see Gispert, Lagache \& Puget 2000). Hence
the CIRB reflects the bulk of the star formation that took place over
the history of the universe. By resolving it into individual galaxies,
we would therefore pinpoint the times and places where most stars seen
in the local universe were formed.  Two physical processes were
considered for its origin: nucleosynthesis, i.e. stellar radiation in
star forming galaxies, and accretion around a black hole, i.e. active
galactic nuclei. In both cases, the light is not directly coming from
its physical source but is reprocessed by dust, i.e. absorbed and
re-radiated thermally by the ``warm'' dust. Both processes are
probably related (see Genzel et al. 1998), but energetic
considerations, based on the presence of massive black holes and on
the amount of heavy elements in local galaxies, suggest that star
formation should by far dominate in the CIRB over AGN activity (Madau
\& Pozzetti 2000, Franceschini et al. 2001). However, until the
individual galaxies responsible for the CIRB are found and studied in
detail, this result will remain theoretical.

The spectral energy distribution (SED) in the IR of local galaxies
peaks above $\sim$60\,$\mu$m and typically around 80 $\pm$ 20\,$\mu$m
(see Sanders \& Mirabel 1996). As a result, the distant galaxies
responsible for the peak of the CIRB detected by COBE around
$\lambda_{\rm max}$ $\sim$ 140\,$\mu$m should be located below $z\sim$
1.3 and present a redshift distribution peaked around $z\sim$ 0.8, if
their SEDs do not strongly differ from those of local galaxies. As we
will see this is also the redshift range of the galaxies detected at
15\,$\mu$m with ISOCAM.

The ISOCAM extragalactic surveys were performed with two filters, LW2
(5-8.5\,$\mu$m) and LW3 (12-18\,$\mu$m), centered at 6.75 and
15\,$\mu$m respectively. The 6.75\,$\mu$m sample of sources is
strongly contaminated by galactic stars, whereas stars are rather
easily distinguished from galaxies at 15\,$\mu$m using optical-MIR
colour-colour plots. As a consequence, we are only concerned here by
the 15\,$\mu$m galaxies. Moreover, the observed 6.75\,$\mu$m light is
no more produced by dust emission for galaxies more distant than
$z\sim$ 0.4 because of k-correction (redshifted stellar light
dominates the 6.75\,$\mu$m band above this redshift), whereas the
observed 15\,$\mu$m light is mostly due to dust emission for galaxies
up to $z\sim$ 2.

About 1000 galaxies detected in the 15\,$\mu$m surveys were used to
produce number counts (i.e. surface density of galaxies as a function
of flux density; see Elbaz et al. 1999). The steep slope of the
15\,$\mu$m counts below $\sim\,1$ mJy indicates the presence of an
excess of faint sources by one order of magnitude in comparison with
predictions assuming no evolution of the 15\,$\mu$m luminosity
function with redshift. The presence of broad emission features in the
MIR spectrum of galaxies alone cannot explain the shape of the number
counts and a strong evolution of either the whole luminosity function
(Xu 2000, Chary \& Elbaz 2001) or preferentially of a sub-population
of starburst galaxies evolving both in luminosity and density
(Franceschini et al. 2001, Chary \& Elbaz 2001, Xu et al. 2001) is
required in order to fit the ISOCAM 15\,$\mu$m counts. In the present
paper, we suggest that these ISOCAM galaxies are in fact dusty
starbursts responsible for the bulk of the CIRB.

In Sect.~\ref{resolve}, we compare the sensitivity of different
extragalactic surveys in several wavelength ranges to detect the
galaxies responsible for the CIRB. It is suggested that MIR is
presently the most efficient technique to detect dusty starbursts up
to $z\sim$ 1.3.

In Sect.~\ref{ebl15}, we calculate the 15\,$\mu$m integrated galaxy
light (IGL) due to ISOCAM galaxies. The 15\,$\mu$m IGL is the sum of
the 15\,$\mu$m fluxes from individual galaxies, down to a given
sensitivity limit, per unit area. It represents a lower limit to the
15\,$\mu$m EBL, which remains unknown. Once the redshift distribution
and SED of these galaxies is determined, it becomes possible to
estimate their contribution to the CIRB.

In Sect.~\ref{MIRtracer}, we demonstrate that MIR luminosities at
6.75, 12 and 15\,$\mu$m are strongly correlated with the bolometric IR
luminosity (from 8 to 1000\,$\mu$m) for local galaxies. The
correlations presented in Chary \& Elbaz (2001) are confirmed here
with a larger sample of galaxies.

Before spectroscopic redshifts are obtained for the full sample of
ISOCAM galaxies used to produce these number counts, the redshift
distribution of these galaxies can be inferred from a few sub-samples:
HDFN (Aussel et al. 1999, 2001), CFRS-14 (Flores et al. 1999, 2002),
CFRS-03 (Flores et al. 2002). The ultra-deep ISOCAM survey of the HDFN
samples a flux density range where most of the evolution observed in
the number counts takes place and where the bulk of the 15\,$\mu$m IGL
is produced. This field is complete in spectroscopic redshifts, so it
is used to estimate the bolometric IR luminosities and star formation
rates of the ISOCAM galaxies in Sect.~\ref{bolIR}.

This result relies on two assumptions:\\
- that the main source for the MIR light in ISOCAM galaxies is
star formation and not accretion around a black hole.\\
- that the correlations found in the local universe between the
MIR and bolometric IR luminosity of galaxies remain valid up to
$z\sim$ 1. 

The first assumption is discussed and justified in
Sect.~\ref{agnfrac}, where soft and hard X-ray data from the Chandra
and XMM-Newton X-ray observatories are combined with ISOCAM data on
galaxies in the HDFN and Lockman Hole regions respectively.

The issue of the robustness of the MIR-FIR correlations in the distant
universe is addressed in Sect.~\ref{radio}, where IR luminosities are
also computed from radio (1.4 GHz) flux densities for a sub-sample of
24 distant and 109 local ISOCAM galaxies.

In Sect.~\ref{ldens}, we compute the cosmic density of IR light due to
luminous IR galaxies ($L_{\rm IR}\geq 10^{11}~L_{\sun}$) at ${\bar
z}\sim$ 1. In Sect.~\ref{cirb}, we evaluate the contribution of the
ISOCAM galaxies to the CIRB, more precisely to its peak emission
around $\lambda_{max}\sim$ 140\,$\mu$m. Finally, the nature of ISOCAM
galaxies is discussed in the conclusions (Sect.~\ref{conclusions}).

In the following, we will use the terms ULIG for galaxies with an IR
luminosity $L_{\rm IR}=~L[8-1000\,\mu{\rm m}]\geq 10^{12}~L_{\sun}$,
LIG, when $10^{11}\leq (L_{\rm IR}/L_{\sun})< 10^{12}$ and luminous IR
galaxies for both ($L_{\rm IR}\geq~10^{11}~L_{\sun}$).  Throughout
this paper, we will assume H$_o$= 75 km s$^{-1}$ Mpc$^{-1}$,
$\Omega_{\rm matter}$= 0.3 and $\Omega_{\Lambda}= 0.7$.
\section{How to resolve the CIRB into individual galaxies ?}
\label{resolve}
\begin{figure*}
\resizebox{\hsize}{!}{\includegraphics{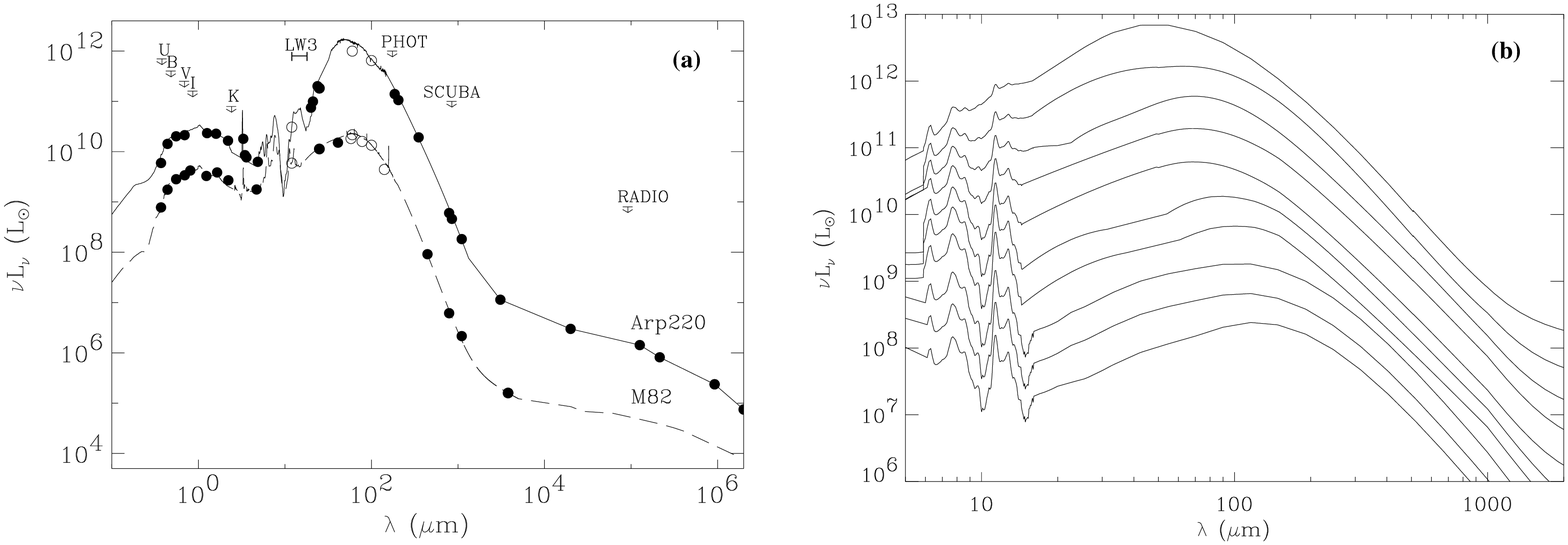}}
\caption{Spectral energy distribution of galaxies in the infrared.
{\bf a)} SEDs of M 82 (dashed line) and Arp 220 (plain line) combining
photometric (filled and open dots) and spectroscopic data. Open dots
are used when they overlap with spectroscopic data. Radio data for Arp 220
from Anantharamaiah et al. (2000). Sub-millimetric data for M 82 from
SCUBA (Hughes et al. 1994). $\lambda=$ 45-190\,$\mu$m: ISO-LWS spectra
from Fischer et al. (1999). Arp220's 18\,$\mu$m silicate absorption
feature from Smith, Aitken \& Roche (1989). $\lambda=$ 5-18\,$\mu$m:
ISOCAM Circular Variable Filter (CVF) spectra (Arp 220: Charmandaris
et al. 1999, M 82: F\"orster Schreiber et al. 2000) normalized to the
IRAS 12\,$\mu$m flux. For M 82, the CVF spectrum of the central region
of the galaxy was complemented outward by adding to it a typical
spectrum of a reflection nebula (NGC 7023, Cesarsky et al. 1996b), in
order to match the 12\,$\mu$m IRAS flux. $\lambda=$2.5 to 4\,$\mu$m:
for M 82, SWS spectrum of F\"orster Schreiber et al. (2001) scaled
using the NGC7023 spectrum of Moutou et al.  (2000); for Arp 220, fit
to the ISOPHOT data (Klaas et al. 1997) combined with the profile from
Moutou et al. (2000). Optical to near-infrared: mix of the Coleman, Wu
and Weedman (1980) templates to match the observed total magnitudes
obtained from NED (Arp 220: 67\,$\%$ Sbc and 33\,$\%$ Ell; M 82:
12\,$\%$ Sbc and 88\,$\%$ Ell). {\bf b)} 10 SEDs with
$log_{10}\left(L_{\rm IR}/L_{\sun}\right)$= 8.5 to 13, with a step of
0.5, from the library of 105 SEDs of Chary \& Elbaz (2001).}
\label{FIG:sedsm82a220}
\end{figure*}
\subsection{Contribution of IRAS galaxies}
One of the major results from the IRAS satellite was to demonstrate
that the bolometric luminosity of galaxies more luminous than $L_{\rm
bol}\sim~10^{11}~L_{\sun}$ was underestimated by more than one order
of magnitude before the IR luminosity was accounted for (Soifer et
al. 1987). These luminous IR galaxies are not typical of the local
galaxy population, since local galaxies radiate only about 30\,$\%$ of
their bolometric luminosity in the dust regime from 8 to 1000\,$\mu$m
and only $\sim$ 2\,$\%$ of the local bolometric luminosity density is
due to luminous IR galaxies. If we convert the IR luminosity into a
star formation rate (SFR) using the formula of Kennicutt (1998),
\begin{equation}
{\rm SFR(M_{\sun}/yr)}=~1.71\times10^{-10}~L_{\rm IR}[8-1000\,\mu {\rm m}] (L_{\sun})
\label{eqSFR}
\end{equation}
we find that luminous IR galaxies ($L_{\rm bol}\sim L_{\rm
IR}\geq~10^{11}~L_{\sun}$) form stars at a rate larger than
$\sim$\,20\,$M_{\sun}$ yr$^{-1}$.  Eq.~(\ref{eqSFR}) assumes
continuous bursts lasting 10-100 Myr, solar abundance and a Salpeter
IMF (Kennicutt 1998).

IRAS surveyed about 95\,$\%$ of the sky down to a completeness limit
of $\sim$ 0.5 mJy at 60\,$\mu$m ($\sim$ 1.5 Jy at 100\,$\mu$m) and the
differential counts are well fitted by an Euclidean slope,
$dN/dS_{\nu}\sim~S_{\nu}^{-2.5}$ (Soifer et al. 1987). The fluxes of
all galaxies detected down to this sensitivity limit by IRAS add up to
a 60\,$\mu$m IGL of $\sim$ 0.15 nW m$^{-2}$ sr$^{-1}$. This is less
than 1\,$\%$ of the value of the CIRB measured by COBE at
$\lambda_{max}\sim$ 140\,$\mu$m, $IGL_{140}$= (25$\pm$7) nW m$^{-2}$
sr$^{-1}$.
\subsection{Contribution of ISOPHOT galaxies}
Deeper FIR extragalactic surveys were performed with ISOPHOT (Lemke et
al. 1996) onboard ISO at 170\,$\mu$m. At this depth, the differential
counts are no more fitted by an Euclidean slope, expected in the case
of no evolution of galaxies with redshift, but instead a strong excess
of faint sources was found. At the confusion limit of ISOPHOT of about
120 mJy, less than 10\,$\%$ of the value of the CIRB at 170\,$\mu$m is
resolved into individual galaxies (Dole et al. 2001). A preliminary
follow-up of the ISOPHOT galaxies suggests that a large fraction of
these galaxies are located at low redshift. 
\begin{figure*}
\resizebox{\hsize}{!}{\includegraphics{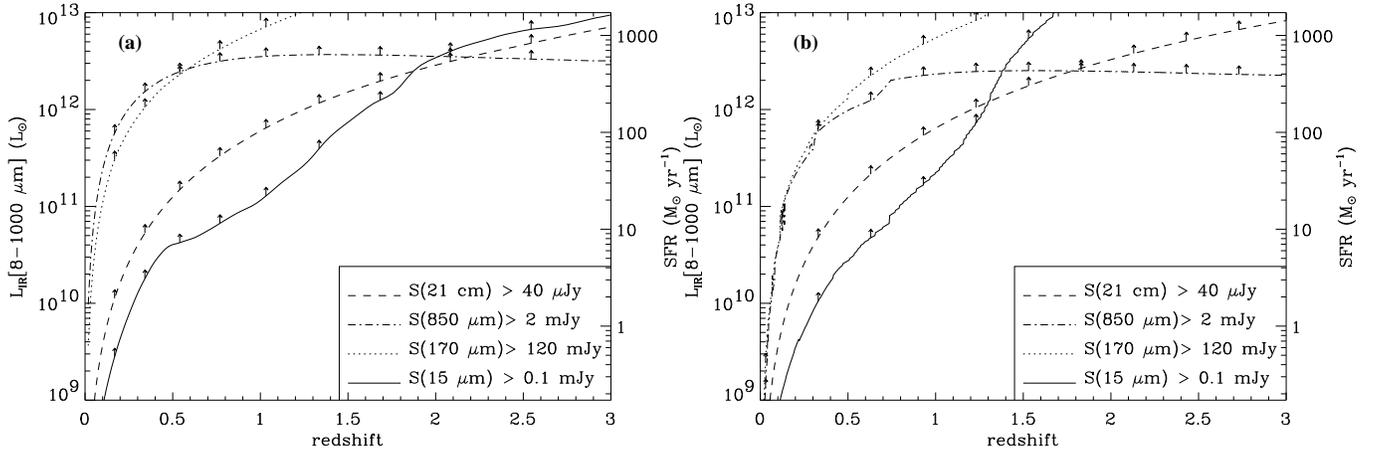}}
\caption{IR luminosity (left axis) and SFR (right axis) corresponding
to the sensitivity limits of ISOCAM (15\,$\mu$m, plain line), ISOPHOT
(170\,$\mu$m, dotted line), SCUBA (850\,$\mu$m, dot-dashed line) and
VLA/WSRT (21 cm, dashed line) as a function of redshift. K-correction
from: {\bf a)} the template SED of M 82
(Fig.~\ref{FIG:sedsm82a220}a), normalized to the appropriate IR
luminosity, {\bf b)} the library of template SEDs from Chary \& Elbaz
(2001), plotted in the Fig.~\ref{FIG:sedsm82a220}b.}
\label{FIG:sensM82}
\end{figure*}
\subsection{Contribution of SCUBA galaxies}
In the sub-millimeter range, the SCUBA bolometer array on the James
Clerk Maxwell Telescope (JCMT) has produced deep images at 450 and
850\,$\mu$m. Number counts at both wavelengths present a steep slope
compatible with a strong evolution as compared to the local universe
(see Smail et al. 2001 and references therein). 

At 450\,$\mu$m, a depth of 10 mJy is reached and the combined fluxes
of all SCUBA galaxies produce about 15\,$\%$ of the CIRB measured by
COBE-FIRAS at this wavelength (Smail et al. 2001).

At 850\,$\mu$m, SCUBA is confusion limited at $\sim$ 2 mJy (Hughes et
al. 1998, Barger, Cowie \& Sanders 1999, Eales et al. 2000, Smail et
al. 2001), because of its large point spread function (PSF) of
15\,$\arcsec$ full width half maximum (FWHM). About 20\,$\%$ of the
value of the CIRB measured by COBE-FIRAS at 850\,$\mu$m (the
850\,$\mu$m EBL) is resolved into galaxies at this depth. However,
using gravitational lensing this limit can be lowered to $\sim$ 1 mJy,
where $\sim$~60\,$\%$ of the CIRB is resolved (Smail et al. 2001,
Blain et al. 1999a).

The 850\,$\mu$m EBL measured by COBE-FIRAS ($\sim$ 0.5$\pm$0.2 nW
m$^{-2}$ sr$^{-1}$) is 50 times lower than the peak value of the CIRB
at $\lambda_{max}\sim$ 140\,$\mu$m. In order to compute the
contribution of SCUBA galaxies to the peak of the CIRB it is therefore
necessary to determine their redshift distribution and SED. Until now,
very few redshifts have been obtained due to the large PSF of SCUBA
and optical faintness of these galaxies. Hence the fraction of the
CIRB resolved into individual galaxies by SCUBA remains highly
uncertain.

\subsection{Comparison with the MIR and radio}
\label{sensrad}
The deepest ISOCAM surveys reach a completeness limit of $\sim$
100\,$\mu$Jy at 15\,$\mu$m (without lensing, but this limit goes down
to $\sim$ 30-50\,$\mu$Jy including lensing, see Sect.~\ref{ebl15}).
For a given galaxy SED and redshift, this flux density can be
converted into a bolometric IR luminosity ($L_{\rm
IR}$[8-1000\,$\mu$m]). Any galaxy more luminous than this $L_{\rm IR}$
will be detected in the survey. In order to compare the sensitivity to
distant luminous IR galaxies of ISOCAM, ISOPHOT and SCUBA, we have
calculated $L_{\rm IR}$ corresponding to their sensitivity limits for
a given redshift between $z$=0 and 3. On one hand, we used the SED of
M 82 (Fig.~\ref{FIG:sedsm82a220}a) for galaxies of all
luminosities. On the other hand, we used a library of template SEDs
from Chary \& Elbaz (2001). In the following, we will call it the
``multi-template'' technique. This library of 105 SEDs was constructed
under the constraint that it reproduces the correlations observed
between the 6.75, 12, 15, 60, 100 and 850\,$\mu$m luminosities of
local galaxies. The SEDs were interpolated from a sample of observed
SEDs, including MIR spectra obtained with the ISOCAM CVF. They cover
the luminosity range $log_{10}\left(L_{\rm IR}/L_{\sun}\right)$= 8.5
to 13.5 (see Fig.~\ref{FIG:sedsm82a220}b).

We did not use Arp 220's SED in Fig.~\ref{FIG:sedsm82a220}a, because
this galaxy presents a large FIR over MIR luminosity ratio, which is
not representative of galaxies within the same luminosity range (see
Figs.\ref{FIG:correl}c,d).

The radio continuum ($\lambda$= 21 cm, $\nu$= 1.4 GHz) is also a
tracer of $L_{\rm IR}$ because of the tight correlation between both
luminosities in the local universe (see Yun, Reddy \& Condon 2001 and
references therein). If this correlation remains valid in the distant
universe then we can translate the sensitivity limit of the deepest
radio surveys into a minimum $L_{\rm IR}$ accessible at a given
redshift assuming a radio SED. The correlation between the radio and
FIR luminosities is usually described by the ``$q$'' parameter (Condon
et al. 1991):
\begin{equation}
q = log_{10}\left( \frac{L_{\rm FIR}({\rm W})}{3.75\times10^{12} ({\rm Hz})} \times \frac{1}{L_{\rm1.4 GHz}({\rm W~Hz}^{-1})}\right)
\label{eq:qpar}
\end{equation}
where $L_{\rm1.4 GHz}$ is the monochromatic luminosity at 1.4 GHz and
$L_{\rm FIR}$ is the FIR luminosity between 40 and 120\,$\mu$m, as
defined by Helou et al. (1988),
\begin{equation}
L_{\rm FIR}= 1.26\times10^{-14}~(2.58~S_{60}+S_{100})\times4\pi d(m)^2
\end{equation}
where $S_{60}$ and $S_{100}$ are the IRAS flux densities at 60 and
100\,$\mu$m in Jy. The relationship between $L_{\rm
FIR}$[40-120\,$\mu$m] and $L_{\rm IR}$[8-1000\,$\mu$m] was computed
with the sample of 300 galaxies from the Bright Galaxy Sample (BGS,
Soifer et al. 1987) detected in the four IRAS bands (12, 25, 60 and
100\,$\mu$m):
\begin{equation}
L_{\rm IR} = (1.91\pm0.17) \times L_{\rm FIR}
\label{eq:irfir}
\end{equation}
$L_{\rm IR}$ is defined as (Sanders \& Mirabel 1996):
\begin{equation}
\begin{tabular}{lll}
$L_{\rm IR}$ & = & $[13.48~S_{12} + 5.16~S_{25} + 2.58~S_{60} + S_{100}]$\\
	   &   & $\times\,1.8\times10^{-14}~\times4\pi d$(m)$^2$
\label{eq:ir}
\end{tabular}
\end{equation}
where $S_{12}$ and $S_{25}$ are the IRAS flux densities at 12 and
25\,$\mu$m in Jy.  Yun, Reddy \& Condon (2001) measured a value of
$q$= 2.34$\pm$0.01 from a flux limited ($S_{60}\geq$ 2 Jy) sample of
1809 IRAS galaxies.

The deepest radio surveys presently available were performed in the
HDFN by Richards (2000) and Garrett et al. (2000) with the VLA and
WSRT respectively. Both surveys reach the same sensitivity limit of
40\,$\mu$Jy at 1.4 GHz (5-$\sigma$). In order to convert this flux
density into a $L_{\rm 1.4 GHz}$ for a given redshift, we used a power
index of $\alpha$= 0.8$\pm$0.15 ($S_{\nu}\propto \nu^{-\alpha}$), as
suggested by Yun, Reddy \& Condon (2001) for starburst galaxies. 

Fig.~\ref{FIG:sensM82} presents the sensitivity of the deepest MIR,
FIR, sub-millimeter and radio surveys in the form of L$_{\rm IR}$ (or
SFR) as a function of redshift. The sensitivity limits used for
ISOCAM, ISOPHOT and SCUBA are taken from the deepest existing surveys
at these wavelengths, as previously described. Fig.~\ref{FIG:sensM82}a
(with M 82) and Fig.~\ref{FIG:sensM82}b (``multi-template'' technique)
both clearly show that the faintest IR galaxies are best detected at
15\,$\mu$m up to $z\sim$ 1.3. The right axis of the plots shows the
corresponding minimum SFR that a galaxy must harbor to be detected at
a given redshift in the surveys. At $z\sim$ 1, ISOCAM (plain line)
detects all LIGs while SCUBA (dot-dashed line) detects only ULIGs. The
difference in sensitivity between ISOCAM and the deepest radio surveys
is about a factor 2. ISOPHOT (dotted line) is only sensitive to ULIGs
above $z\sim$ 0.5. In Fig.~\ref{FIG:sensM82}b, where we used the
library of template SEDs from Chary \& Elbaz (2001), the plain line
corresponding to ISOCAM rises faster above $z\sim$ 1 than in
Fig.~\ref{FIG:sensM82}a, where we used M 82's SED.  This different
behavior comes from the fact that at these redshifts ISOCAM measures
fluxes at about 7\,$\mu$m in the rest-frame of the galaxy and we will
show that $L_{\rm IR}$/$L_{\rm 7\,\mu m}$ increases with $L_{\rm IR}$
(Sect.~\ref{MIRtracer}).

Above z$\sim$2, the sub-millimeter becomes the most efficient
technique, although only galaxies more luminous than a few 10$^{12}$
L$_{\sun}$ can be detected above a sensitivity limit of 2 mJy at
850\,$\mu$m. Hence the unlensed ISOCAM and SCUBA surveys are not
sampling the same redshift and luminosity ranges.  This statement is
confirmed observationally in the HDFN itself where none of the ISOCAM
sources are detected by SCUBA (Hughes et al.  1998) and over a larger
scale in the Canada France Redshift Survey 14 (CFRS-14, Eales et
al. 2000). In this latter field of $\sim$50$\arcmin ^2$, only the two
brightest 15\,$\mu$m ISOCAM sources are detected at 850\,$\mu$m, among
a sample of 50 ISOCAM (Flores et al. 1999) and 19 SCUBA sources. This
confirms that both instruments detect different sets of objects. More
common objects between ISOCAM and SCUBA are expected if they are
gravitationally lensed, since the sensitivity limits of both
instruments are then decreased by a factor of about two. Indeed in the
clusters A 370 and A 2390, three of the four lensed SCUBA galaxies
(Smail et al. 2001) are also detected with ISOCAM (Altieri et
al. 1999, Metcalfe 2000).

\section{The 15\,$\mu$m extragalactic background light}
\label{ebl15}
\begin{figure*}
\resizebox{\hsize}{!}{\includegraphics{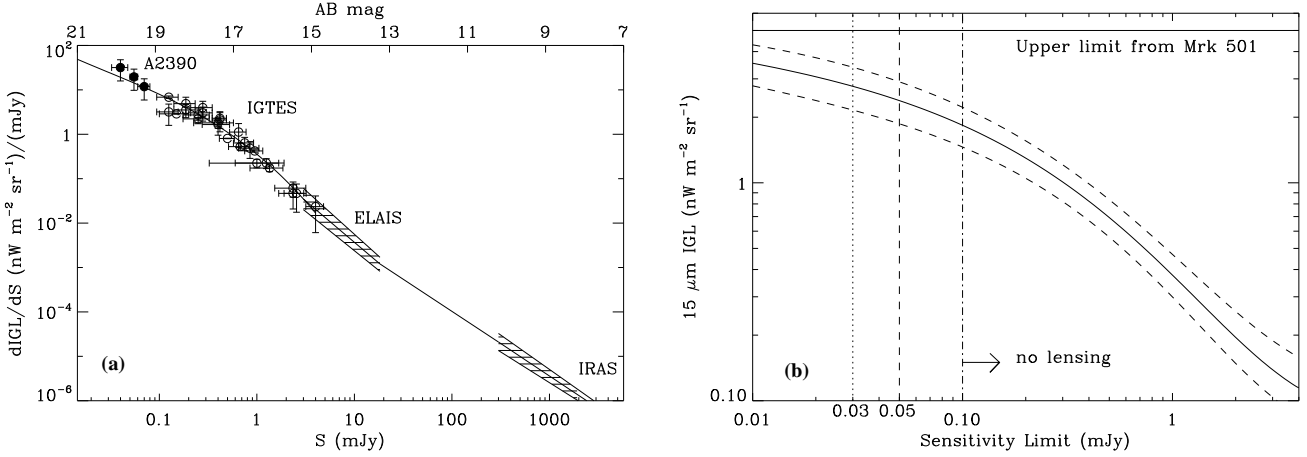}}
\caption{{\bf a)} Differential contribution to the 15\,$\mu$m
Integrated Galaxy Light as a function of flux density and AB
magnitude. The plain line is a fit to the data: Abell 2390 (Altieri et
al. 1999), the ISOCAM Guaranteed Time Extragalactic Surveys (IGTES,
Elbaz et al. 1999), the European Large Area Infrared Survey (ELAIS,
Serjeant et al. 2000) and the IRAS all sky survey (Rush, Malkan \&
Spinoglio 1993). {\bf b)} Contribution of ISOCAM galaxies to the
15\,$\mu$m extragalactic background light (EBL), i.e. 15\,$\mu$m
Integrated Galaxy Light (IGL), as a function of sensitivity or $AB$
magnitude $\left(AB=\ -2.5\,log\left(S_{\rm
mJy}\right)+16.4\right)$. The plain line is the integral of the fit to
dIGL/dS (Fig.\ref{FIG:dEBL}a).  The dashed lines correspond to
1-$\sigma$ error bars obtained by fitting the 1-$\sigma$ upper and
lower limits of dIGL/dS.}
\label{FIG:dEBL}
\end{figure*}
Above the Earth's atmosphere, the 15\,$\mu$m light is strongly
dominated by the zodiacal emission from interplanetary dust and it has
not yet been possible to make a direct measurement of the 15\,$\mu$m
background, or EBL. Individual galaxies contribute to this background
and a lower limit to the 15\,$\mu$m EBL can be obtained by adding up
the fluxes of all ISOCAM galaxies detected per unit area down to a
given flux limit. The resulting value is called the 15\,$\mu$m
integrated galaxy light (IGL).

The statistical reliability with which the surface density of MIR
sources is computed depends on their flux: large and shallow fields
are required for bright sources which are less numerous, whereas faint
sources require smaller and deeper fields. The optimization is
obtained by observing similar volumes of the sky but in pencil beams
for faint sources and in wide surveys for bright sources, instead of
one large and deep survey which would be unnecessarily time
consuming. Together with the need to account for cosmic variance, this
explains why we have combined a set of 10 surveys located in 7 regions
of the northern and southern hemispheres, in order to estimate the
15\,$\mu$m IGL. These surveys cover five orders of magnitude in flux
density from $\sim50$ $\mu$Jy to $\sim4$ Jy (including IRAS counts at
12\,$\mu$m). The number counts presented in Elbaz et al. (1999) are
converted here into a differential contribution to the 15\,$\mu$m IGL
as a function of flux density, estimated from the following formula:
\begin{equation}
\frac{dIGL}{dS}=~\frac{dN}{dS}\times~\left(\frac{S_{15}}{10^{20}}\right)~\times~\nu_{15}
\end{equation}
where $dN$(sr$^{-1}$) is the surface density of sources with a flux
density $S_{\nu}$[15\,$\mu$m]$=S_{15}$ (mJy) over a flux density bin
$dS$(mJy) (1 mJy= 10$^{-20}$ nW m$^{-2}$ Hz$^{-1}$) and $\nu_{15}$(Hz)
is the frequency of the 15\,$\mu$m photons.

Below $S_{15}\sim$ 3 mJy, about 600 galaxies were used to produce the
points with errors bars in Fig.~\ref{FIG:dEBL}a. Fig.~\ref{FIG:dEBL}b
shows the 15\,$\mu$m IGL as a function of depth. It corresponds to the
integral of Fig.~\ref{FIG:dEBL}a, where the data below 3 mJy were
fitted with a polynomial of degree 3 and the 1-$\sigma$ error bars on
$dIGL/dS$ were obtained from the polynomial fit to the upper and lower
limits of the data points. The 15\,$\mu$m IGL does not converge above
a sensitivity limit of $S_{15}\sim$ 50\,$\mu$Jy, but the flattening of
the curve below $S_{15}\sim 0.4$ mJy suggests that most of the
15\,$\mu$m EBL should arise from the galaxies already unveiled by
ISOCAM.

Above the completeness limit of $S_{15}\sim 50~\mu$Jy, we computed
an IGL of:
\begin{equation}
IGL_{15}(S_{15}\geq 50\,\mu {\rm Jy}) = 2.4\pm0.5~{\rm nW~m^{-2}~sr^{-1}}
\label{eq:igl}
\end{equation}
where the error bar corresponds to the 68\,$\%$ confidence level
(i.e. 1-$\sigma$). The error bar combines the uncertainty on the
measurements for each individual survey with cosmic variance. Each
flux bin is covered by two to three independent surveys. ISOCAM number
counts are only complete above $S_{15}\sim$ 100\,$\mu$Jy, but Aussel
et al. (1999) computed number counts down to $S_{15}\sim 50~\mu$Jy
after correcting for incompleteness and a consistent result is
obtained by Altieri et al. (1999) after correcting for lensing
magnification a sample of galaxies in the field of the galaxy cluster
Abell 2390. 

Only 9 sources are detected below $S_{15}\sim 50~\mu$Jy by Altieri et
al. (1999), where the flux determination and correction for
completeness are less robust and are not confirmed by another
survey. A tentative value of $IGL_{15}\sim$ 3.3 $\pm$ 1.3 nW m$^{-2}$
sr$^{-1}$ was quoted in Altieri et al. (1999), above $S_{15}$=
30\,$\mu$Jy. This value may be slightly overestimated if we consider
that the models of Franceschini et al. (2001) and Chary \& Elbaz
(2001), which reproduce the number counts from ISOCAM at 15\,$\mu$m,
from ISOPHOT at 90 and 170\,$\mu$m and from SCUBA at 850\,$\mu$m, as
well as the shape of the CIRB from 100 to 1000\,$\mu$m, consistently
predict a 15\,$\mu$m EBL of:
\begin{equation}
EBL^{models}(15\,\mu {\rm m}) \sim 3.3~{\rm nW~m^{-2}~sr^{-1}}
\end{equation}

If this prediction from the models is correct then about $73\pm15\,\%$
of the 15\,$\mu$m EBL is resolved into individual galaxies by the
ISOCAM surveys. 

This result is consistent with the upper limit on the 15\,$\mu$m EBL
estimated by Stanev \& Franceschini (1998) of:
\begin{equation}
EBL^{max}(15\,\mu {\rm m}) \sim 5~{\rm nW~m^{-2}~sr^{-1}}
\end{equation}
This upper limit was computed from the 1997 $\gamma$-ray outburst of
the blazar Mkn 501 ($z=$ 0.034) as a result of the opacity of MIR
photons to $\gamma$-ray photons, which annihilate with them through
$e^+e^-$ pair production. It was recently confirmed by Renault et
al. (2001), who found an upper limit of 4.7 nW m$^{-2}$ sr$^{-1}$ from
5 to 15\,$\mu$m.

Finally, the 15\,$\mu$m background (EBL) must be contained between
the following limits:
\begin{equation}
2.4\pm0.5 \leq \frac{EBL(15\,\mu {\rm m})}{\rm \left(nW~m^{-2}~sr^{-1}\right)}\leq 5
\end{equation}

\begin{table} 
\begin{tabular}{|l|l|l|l|}
\hline
Field         & F(mJy)        & $\%$ & Ref. \\
\hline
A2390,HDFN    & [ 0.05, 0.1 ] & 23   & (1,2)  \\
IGTES         & [ 0.1, 0.5 ]  & 48   & (2,3)  \\
IGTES         & [ 0.5, 3 ]    & 23   & (3)  \\
ELAIS         & [ 3, 32 ]     &  4.5 & (4)  \\
Interpolation & [ 32,300 ]    &  1   & (5)  \\
IRAS          & [ 300,4000 ]  &  0.5 & (6)  \\
\hline
\end{tabular}
\caption{Relative contributions to the 15\,$\mu$m IGL as a function of
flux density. Ref.: (1) Altieri et al. (1999), Metcalfe (2000), (2)
Aussel et al. (1999, 2001), (3) Elbaz et al. (1999), (4) Serjeant et
al. (2000), renormalized by a factor 0.5 as in Genzel \& Cesarsky
(2000), following suggestion from S.Serjeant for photometric
calibration, (5) Interpolation in a flux density range where we do not
have enough statistics. We used the same slope as in the IRAS counts,
(6) IRAS counts from Rush, Malkan \& Spinoglio (1993) at 12\,$\mu$m.}
\label{TAB:frac}
\end{table}

In Table~\ref{TAB:frac}, we give the relative contribution to the
15\,$\mu$m IGL from sources within a given flux density range.  Nearly
all the IGL is produced by sources fainter than 3 mJy (94\,$\%$) and
about 70\,$\%$ by sources fainter than 0.5 mJy. This means that the
nature and redshift distribution of the galaxies producing the bulk of
the 15\,$\mu$m IGL can be determined by studying these faint galaxies
only. We will therefore assume in the following that the
spectroscopically complete sample of 15\,$\mu$m galaxies detected in
the HDFN below 0.5 mJy (Aussel et al. 1999, 2001) is representative of
the population of dusty galaxies producing the bulk of the 15\,$\mu$m
IGL.

\section{MIR, as a tracer of the bolometric IR luminosity}
\label{MIRtracer}
\subsection{SED of galaxies in the IR}
\begin{figure}
\resizebox{\hsize}{!}{\includegraphics{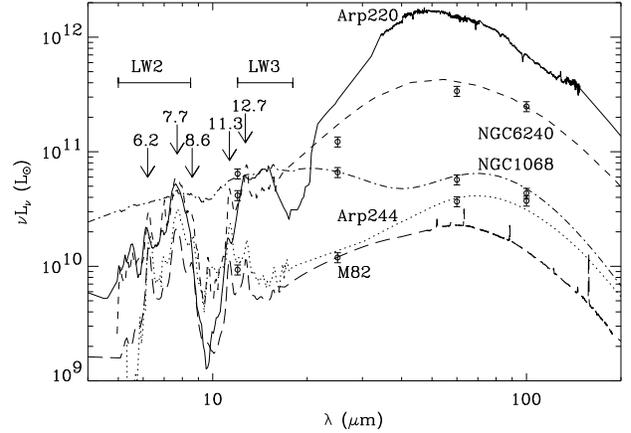}}
\caption{Spectral energy distributions in the infrared (5-200\,$\mu$m)
of 3 starbursts (M 82, Arp 244, Arp 220) and 2 Seyfert 2's (NGC 1068,
NGC 6240). MIR spectra from ISOCAM CVF: NGC 1068 (Laurent et al. 2000,
Le Floc'h et al. 2001), Arp 244 (Vigroux et al. 1996, Mirabel et
al. 1998), NGC 6240 (Charmandaris et al. 1999). FIR: fit of the IRAS
flux densities at 12, 25, 60 and 100\,$\mu$m using the template SEDs
from Chary \& Elbaz (2001) for Arp 244 and NGC 6240. For NGC 1068, the
IRAS flux densities were fitted by the sum of two modifified black
bodies of temperatures 30 and 170 K as in Telesco et al. (1984). Open
dots with error bars: IRAS data (from NED) for Arp244, NGC 6240 and
NGC 1068.}
\label{FIG:sedir}
\end{figure}
Above a bolometric luminosity of a few $10^{10}~L_{\sun}$, the SED of
a galaxy is dominated by dust radiation with respect to direct stellar
emission. Stellar light is absorbed by dust and re-radiated in the IR
regime above $\lambda\sim$ 5\,$\mu$m. The full SEDs, from the UV to
the radio, of the starburst galaxy, M 82, and of the ULIG, Arp 220,
are plotted in the Fig.~\ref{FIG:sedsm82a220}a. A zoom on the
5-2000\,$\mu$m wavelength range of the SEDs is shown in the
Fig.\ref{FIG:sedir}, together with three other galaxies: a starburst,
Arp 244, and two Seyfert 2 galaxies, NGC 1068 and NGC 6240. Arp 244,
also known as the ``Antennae'' galaxy, is a system of two interacting
spiral galaxies (Vigroux et al. 1996, Mirabel et al. 1998). The
optical, MIR and FIR luminosities of these five galaxies are given in
the Table~\ref{TAB:lum}. Note the increase of the MIR luminosity
together with $L_{\rm FIR}$, or $L_{\rm IR}$, from M 82 to Arp 244 to
NGC 6240. This correlation will be statistically established in
Sect.~\ref{form_correl}. 

\begin{figure*}
\resizebox{\hsize}{!}{\includegraphics{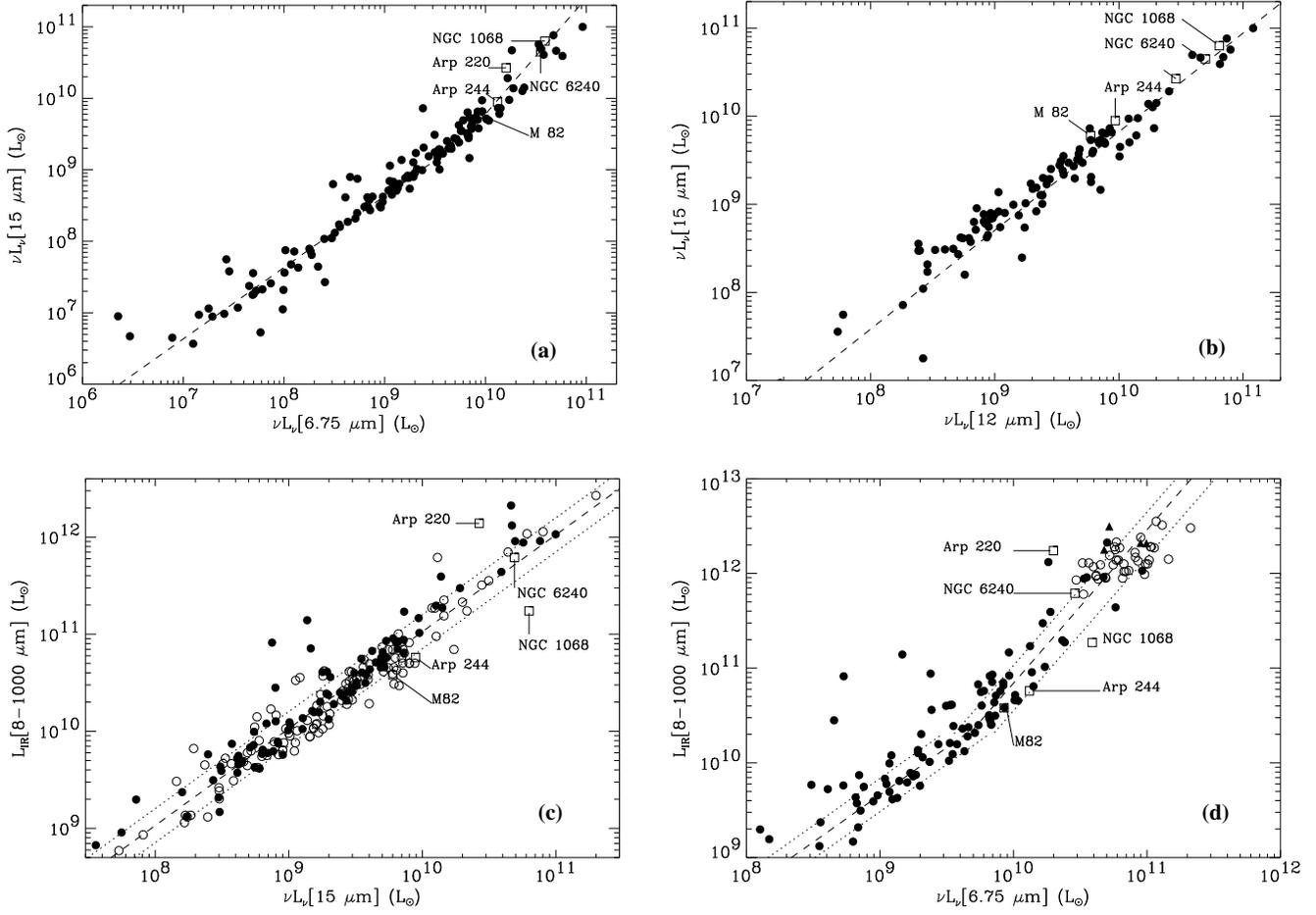}}
\caption{IR luminosity correlations for local galaxies (see text for
the origin of the sample). The five galaxies from Fig.\ref{FIG:sedir}
and Table~\ref{TAB:lum} are marked with open squares. {\bf a)}
ISOCAM-LW3 (15\,$\mu$m) versus ISOCAM-LW2 (6.75\,$\mu$m) luminosities
($\nu L_{\nu}$) (56 galaxies). {\bf b)} ISOCAM-LW3 (15\,$\mu$m) versus
IRAS-12\,$\mu$m luminosities (45 galaxies). {\bf c)} $L_{\rm
IR}$[8-1000\,$\mu$m] versus ISOCAM-LW3 (15\,$\mu$m) luminosity (120
galaxies). {\bf d)} $L_{\rm IR}$[8-1000\,$\mu$m] versus
LW2-6.75\,$\mu$m luminosities (91 galaxies). Filled dots: galaxies from
the ISOCAM guaranteed time (47 galaxies including the open
squares). Open dots: 40 galaxies from Rigopoulou et al. (1999). Empty
triangles: 4 galaxies from Tran et al. (2001). Galaxies below $L_{\rm
IR}\sim 10^{10}~L_{\sun}$ present a flatter slope and have $L_{\rm
IR}$/$L_{B}<$ 1.}
\label{FIG:correl}
\end{figure*}

The FIR over MIR luminosity ratio of Arp 220 is larger than for
galaxies of equivalent luminosity (see Fig.~\ref{FIG:correl} in
Sect.~\ref{form_correl}). The flat FIR over MIR ratio for NGC 1068 is
typical of the hot dust continuum from AGNs. NGC 1068 is the closest
Seyfert 2 ($L_{\rm IR}=~1.7\times10^{11}~L_{\sun}$).  On one hand, its
MIR spectrum is completely dominated by a continuum due to the hot
dust heated by the central AGN, which produces about 75\,$\%$ of the
MIR luminosity (Le Floc'h et al. 2001). On the other hand, the bulk of
its FIR luminosity originates from the diffuse region surrounding the
nucleus as shown by the SCUBA image at 450\,$\mu$m, which is
associated with star formation (Le Floc'h et al. 2001).  As a result,
the FIR over MIR luminosity ratio is much lower for NGC 1068 than for
starbursts (see Fig.\ref{FIG:sedir}). This galaxy is a perfect
illustration of the close connection between star formation and AGN
activity.

\begin{table}[t!]
\begin{tabular}{|l|l|l|l|l|l|}
\hline
                           & M 82  &Arp244 &Arp220 & N1068 & N6240  \\
\hline
$d(Mpc)$                   & 3.25 &  22.0 &  78.4 &  14.4 &  99.6  \\
$log(L_{B})$               & 10.2 &  10.2 &  10.6 &  10.5 &  10.5  \\
$log(L_{\rm IR})$          & 10.6 &  10.8 &  12.2 &  11.2 &  11.8  \\
$log(L_{FIR})$             & 10.4 &  10.7 &  12.1 &  10.8 &  11.7  \\
$log(L_{LW3})$             &  9.8 &  10.0 &  10.6 &  10.8 &  10.7  \\
$log(L_{LW2})$             &  9.9 &  10.1 &  10.3 &  10.6 &  10.5  \\
\hline
$L_{\rm IR}$/$L_{B}$       &  2   &   3   &  39   &   5   &  18    \\
$L_{\rm IR}$/$L_{LW3}$     & 15   &   7   &  41   &   5   &  10    \\
$L_{\rm IR}$/$L_{LW2}$     &  4   &   4   &  87   &   4   &  22    \\
\hline
\end{tabular}
\caption{Luminosities of 3 starbursts (M 82, Arp 244, Arp 220) and 2
Seyfert 2's (NGC 1068, NGC 6240). $L_{B}$ is the B-band monochormatic
luminosity ($\nu L_{\nu}$ at 0.44\,$\mu$m) from RC3 (de Vaucouleurs et
al. 1991). $L_{\rm IR}$ and $L_{FIR}$ are the IR and FIR luminosities
integrated over 8-1000 and 40-500\,$\mu$m respectively from the
spectra shown in the Fig.\ref{FIG:sedir}. $L_{LW3}$ and $L_{LW2}$ are
the monochromatic luminosities ($\nu L_{\nu}$) at the central
wavelengths 15 and 6.75\,$\mu$m of the ISOCAM LW3 and LW2 filters
respectively. At z$\sim$ 1, the LW3 filter (12-18\,$\mu$m) matches the
rest-frame LW2 bandpass (5-8.5\,$\mu$m).}
\label{TAB:lum}
\end{table}

The MIR (5-40\,$\mu$m) and FIR (40-1000\,$\mu$m) emission of a galaxy
combine three major components (see Fig.\ref{FIG:sedir}):

- broad emission features and their associated underlying continuum,
which dominate in the $\lambda\sim$ 5-10\,$\mu$m domain, and up to
$\sim$~20\,$\mu$m for galaxies with moderate star formation. These
bands located at 6.2, 7.7, 8.6, 11.3 and 12.7 $\mu$m, are usually
quoted as PAHs (polycyclic aromatic hydrocarbons, L\'eger $\&$ Puget
1984, Puget $\&$ L\'eger 1989, Allamandola et al. 1989), although
their exact nature remains uncertain, e.g. Jones \& d'Hendecourt
(2000).

- the ``warm'' dust continuum produced by very small dust grains
(VSGs, smaller than 0.01\,$\mu$m, Andriesse 1978, D\'esert, Boulanger
\& Puget 1990) heated at temperatures larger than 200 K without
reaching thermal equilibrium. This component dominates the SED between
10-20 to 40\,$\mu$m for luminous galaxies.

- the ``cold'' dust continuum produced by big dust grains (``big
grains'', of size $>~0.01\,\mu$m) in the thermal equilibrium at cool
temperatures (below 60-70 K typically). This component is responsible
for the bulk of the FIR light where the SED peaks.

\subsection{The MIR-FIR correlation}
\subsubsection{The database}
\label{data}
We have compiled a database containing MIR and FIR fluxes for a sample
of 157 galaxies detected with ISOCAM and IRAS: 50 galaxies from
Boselli et al. (1998), 44 galaxies from Roussel et al. (2001), 19
galaxies from Laurent et al. (1999, 2000), 41 galaxies from Dale et
al. (2001), plus M 82, Arp 244 and NGC 1068 (from
Figs.~\ref{FIG:sedir}). Fig.~\ref{FIG:correl} shows that the 6.75, 12,
15 and bolometric IR (8-1000\,$\mu$m) luminosities ($L_{\rm IR}$) of
these galaxies are tightly correlated with each other. $L_{\rm IR}$ is
calculated following Eq.~(\ref{eq:ir}). At 15\,$\mu$m (LW3 filter), we
completed our sample with the NEPR sample of IRAS galaxies observed
with ISOCAM-LW3 (Aussel et al. 2000). We converted the 60\,$\mu$m
luminosities of the NEPR galaxies into $L_{\rm IR}$ by fitting the IR
versus 60\,$\mu$m correlation obtained with the BGS (Soifer et
al. 1987). In order to complete our sample in the high luminosity
range at 6.75\,$\mu$m (LW2 filter), we used a sample of 40 ULIGs
observed with the spectrometer of ISOPHOT, PHOT-S (from Rigopoulou et
al. 1999: the conversion into LW2 flux densities was computed as in
Chary \& Elbaz 2001). We also used four ULIGs observed with the ISOCAM
CVF from Tran et al. (2001).  We conservatively selected these star
forming galaxies with respect to AGNs from their line over continuum
ratioas suggested in Tran et al. (2001). The ISOCAM-CVF spectra were
converted into LW2 flux densities using the transmission curve of the
LW2 filter.
\subsubsection{Correlations}
\label{form_correl}
The 6.75, 12 and 15\,$\mu$m luminosities of local galaxies are tightly
correlated with each other (see Fig.~\ref{FIG:correl}a,b) following
the formulae:
\begin{equation}
\begin{tabular}{lll}
$\nu L_{\nu}[15\,\mu {\rm m}]$&=&$0.43\times~\nu L_{\nu}[6.75\,\mu {\rm m}]$ \\
&&for $\nu L_{\nu}[6.75\,\mu {\rm m}]< 5\times10^9\,L_{\sun}$ \\
&=&$3.8\times10^{-7}\times~(\nu L_{\nu}[6.75\,\mu {\rm m}])^{1.62}$\\
&&for $\nu L_{\nu}[6.75\,\mu {\rm m}]\geq 5\times10^9\,L_{\sun}$
\label{eq:7mic}
\end{tabular}
\end{equation}
\begin{equation}
\begin{tabular}{lll}
$\nu L_{\nu}[15\,\mu {\rm m}]$&=&$0.042\times~(\nu L_{\nu}[12\,\mu {\rm m}])^{1.12}$
\label{eq:12mic}
\end{tabular}
\end{equation}
These correlations can be used to estimate the rest-frame 15\,$\mu$m
luminosity for galaxies located around $z\sim$ 0.3 and 1. At these
redshifts, 15\,$\mu$m observed corresponds to about 12 and 7\,$\mu$m
in the rest-frame.

The MIR luminosity of galaxies is a good tracer of $L_{\rm IR}$ as 
seen in Figs.~\ref{FIG:correl}c,d, where $L_{\rm IR}$ is correlated
with $\nu L_{\nu}$ at 15 and 6.75\,$\mu$m following the formulae:
\begin{equation}
\begin{tabular}{lll}
$L_{\rm IR}$&=&$11.1^{+5.5}_{-3.7}\times~(\nu L_{\nu}[15\,\mu {\rm m}])^{0.998}$
\end{tabular}
\label{eq:15mic}
\end{equation}
\begin{equation}
\begin{tabular}{lll}
$L_{\rm IR}$&=&$4.78^{+2.37}_{-1.59}\times~(\nu L_{\nu}[6.75\,\mu {\rm m}])^{0.998}$ \\
&&for $\nu L_{\nu}[6.75\,\mu {\rm m}]< 5\times10^9\,L_{\sun}$ \\
&=&$4.37^{+2.35}_{-2.13}\times~10^{-6}\times~(\nu L_{\nu}[6.75\,\mu {\rm m}])^{1.62}$ \\
&&for $\nu L_{\nu}[6.75\,\mu {\rm m}]\geq 5\times10^9\,L_{\sun}$
\end{tabular}
\label{eq:ir7mic}
\end{equation}

The formulae used in these four correlations are consistent with Chary
\& Elbaz (2001), although slightly more data were used here. The
change of slope in Fig.~\ref{FIG:correl}d and Eq.~(\ref{eq:ir7mic})
from a slope of 1 to a slope larger than 1 statistically establishes
what was suggested by Fig.~\ref{FIG:sedir}: $L_{\rm IR}$/$\nu
L_{\nu}$[6.75\,$\mu$m] increases with $L_{\rm IR}$ for galaxies with
$L_{\rm IR}>$ 2$\times$10$^{10}~L_{\sun}$.

The position of the five templates from Fig.~\ref{FIG:sedir} are
indicated with open squares on Fig.~\ref{FIG:correl}. Two galaxies do
not fit the correlations: Arp 220 and NGC 1068. Arp 220 presents a
very large FIR over MIR ratio, which was interpreted as being due to
extinction in the MIR by Haas et al. (2001), while the IR SED of the
Seyfert 2 NGC 1068 is flat, which is typical of AGNs. In this sense
NGC 6240 is quite atypical, since it falls perfectly on the
correlation followed by star forming galaxies. The presence of an
active nucleus in NGC 6240 was revealed by its strong X-ray luminosity
and Ikebe et al. (2000) suggested that a large fraction of the IR
luminosity of this galaxy could originate from its AGN. However, the
presence of strong MIR features (Genzel et al. 1998), the absence of a
strong continuum emission below $\lambda\sim$ 6\,$\mu$m (Laurent et
al. 2000), together with a FIR over MIR ratio typical of star forming
galaxies, suggest that the bulk of its IR luminosity may still be due
to star formation.
\begin{figure*}
\resizebox{\hsize}{!}{\includegraphics{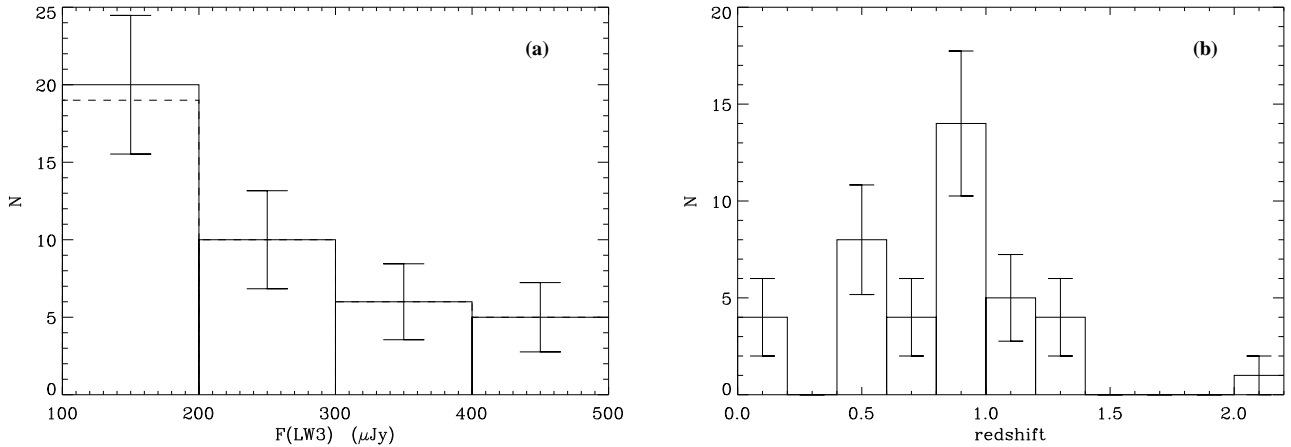}}
\caption{{\bf a)} Histogram of the flux densities of the 41
ISOCAM-HDFN galaxies ($S_{15}\geq$ 0.1 mJy, Aussel et al. 1999, 2001)
with poissonian error bars. Only one galaxy has no spectroscopic
redshift (dashed line). {\bf b)} Redshift distribution of the 40
galaxies with spectroscopic redshift (98\,$\%$ complete in redshift,
poissonian error bars).}
\label{FIG:hisHDF}
\end{figure*}
\section{IR luminosities and star formation rates of ISOCAM galaxies}
\label{bolIR}
In this Section, we compute the IR luminosity and star formation rate
of the galaxies with $S_{15}\geq$ 0.1 mJy located in the Hubble Deep
Field North plus its flanking fields, hereafter called ISOCAM-HDFN
galaxies (Aussel et al. 1999, 2001). This field presents the advantage
of being spectroscopically complete and deep enough to cover a flux
density range ($S_{15}\leq$ 0.5 mJy) where more than 70\,$\%$ of the
15\,$\mu$m IGL is produced. Moreover, the redshift distribution of the
ISOCAM-HDFN galaxies (${\bar z}\sim$ 0.8) is similar to the one
measured in the CFRS fields CFRS-14 (Flores et al. 1999, 2002) and
CFRS-03 (Flores et al. 2002). Together, these fields of 100$\arcmin^2$
each, possess 81 galaxies with $S_{15}\geq$ 0.35 mJy. The redshift
distribution of 70\,$\%$ of the CFRS galaxies, for which a
spectroscopic redshift was measured, is very similar to the one
presented below for the ISOCAM-HDFN galaxies and peaks around $z\sim$
0.7.

\subsection{Redshift distribution of the ISOCAM-HDFN galaxies}
\label{nature}
The list of ISOCAM-HDFN sources used in the following comes from a
revised version of the list published in Aussel et al. (1999). In this
version (Aussel et al. 2001), 4 ISOCAM sources were deblended using
PSF deconvolution based on the position of the optical counterparts
and the flux density of 10 other galaxies were re-evaluated. With a
PSF FWHM of 4.6$\arcsec$ at 15\,$\mu$m (Okumura 1998), optical
counterparts are easily identified for ISOCAM galaxies and all of them
are found within 3$\arcsec$ of the PSF centroid. This can also be seen
is the crowded field of the nearby galaxy cluster Abell 1689 (Fadda et
al. 2000). More than 90\,$\%$ of the ISOCAM galaxies present an
optical counterpart brighter than I(Kron-Cousins)=22.5, which is
natural since ISOCAM is mostly sensitive to galaxies closer than
$z\sim$ 1.3 (see Fig.~\ref{FIG:sensM82}). It results from a
combination of sensitivity limit and k-correction: dust emission
becomes negligible below $\sim 6~\mu$m (except for bright AGNs which
can be detected further away).

A total of 41 ISOCAM-HDFN galaxies have 15\,$\mu$m flux densities
between $S_{15}=$ 0.1 and 0.5 mJy (see Fig.\ref{FIG:hisHDF}a) over a
field of 25.8$'^{2}$ (HDFN+FF). Monte-Carlo simulations give a
completeness of $\sim$~90\,$\%$ in this flux density range (Aussel et
al. 1999, 2001). All but one galaxy (HDF\_PS3\_6a) possess a
spectroscopic redshift (98\,$\%$ complete). Their mean and median
redshift is $\bar{ z} \sim$ 0.8. Except for one galaxy located at
z=2.01 (HDF\_PM3\_5, classified as an AGN, see Sect.~\ref{agnfrac})
and four galaxies closer than z=0.15, all ISOCAM galaxies have a
redshift between $z\sim$ 0.4 and 1.3 (see Fig.\ref{FIG:hisHDF}b), as
expected from the Fig.~\ref{FIG:sensM82}.

\begin{figure*}
\resizebox{\hsize}{!}{\includegraphics{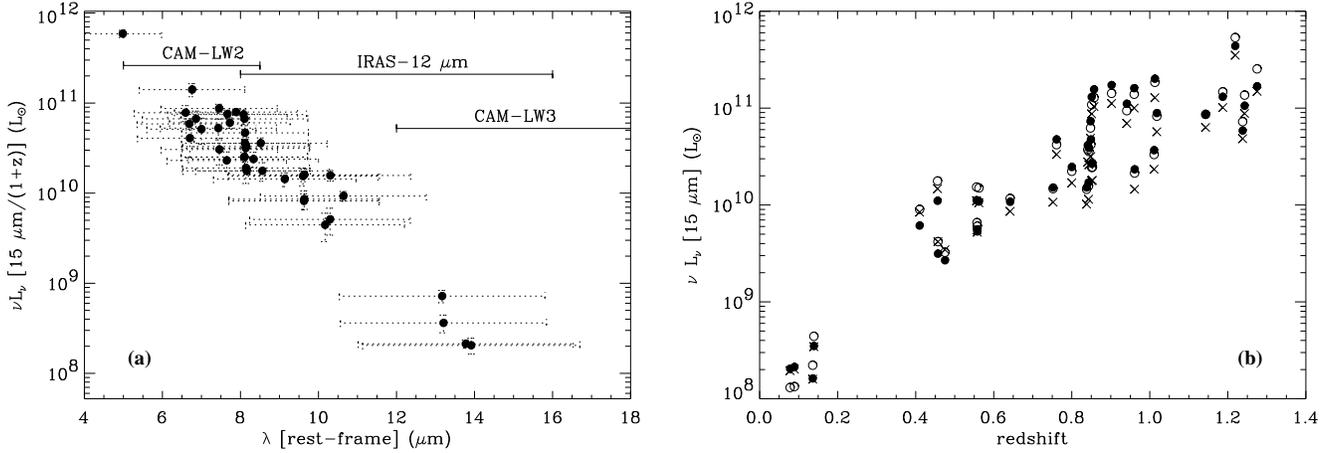}}
\caption{Rest-frame MIR luminosity of ISOCAM-HDFN galaxies. {\bf a)}
luminosity of ISOCAM galaxies at $\lambda_{\rm rf}$= 15/(1+z)\,$\mu$m
as a function of $\lambda_{\rm rf}$. The dotted lines represent the
redshifted LW3 (12-18\,$\mu$m) band. {\bf b)} rest-frame 15\,$\mu$m
luminosity ($\nu L_{\nu}$) versus redshift estimated with three
different techniques: M 82 SED (crosses), ``closest filter'' (filled
dots) and ``multi-template'' (open dots).}
\label{FIG:kcor}
\end{figure*}
\subsection{Fraction of AGN galaxies}
\label{agnfrac}
Cohen et al. (2000) classified the optical spectra of 2/3 of the
ISOCAM galaxies and found only two galaxies presenting broad emission
lines typical of AGNs: 

HDF$\_$PM3$\_$6 (C36367$\_$1346, $z=$ 0.960),

HDF$\_$PM3$\_$20 (C36463$\_$1404, $z=$ 0.962).

However, dusty AGNs may not be easily identified in the optical.

The best technique at present for the identification of dusty AGNs is
offered by hard X-ray observations, which are much less affected by
extinction. The deepest soft plus hard X-ray observation currently
available was performed with the Chandra X-ray Observatory in the HDFN
by Hornschemeier et al. (2001) and Brandt et al. (2001,2002). The
Chandra and ISOCAM catalogs were cross-correlated by Fadda et
al. (2001). A total of 16 ISOCAM-HDFN ($S_{15}\geq$ 0.1 mJy) are
detected in the 1 Msec Chandra image at the depth of $S$[0.5-2
keV]$\simeq$ 3$\times$10$^{-17}$ erg cm$^{-2}$ s$^{-1}$ and $S$[2-8
keV]$\simeq$ 2$\times$10$^{-16}$ erg cm$^{-2}$ s$^{-1}$. They
represent 39\,$\%$ of the ISOCAM-HDFN galaxies. However, the Chandra
image is so deep that it is able to detect starburst galaxies up to
large redshifts and among the 16 ISOCAM galaxies in common with
Chandra sources, only 5 are classified as being AGNs or AGN dominated
by Fadda et al. (2001). These 5 sources include the two galaxies
previously identified as AGNs from their optical spectra and the
following 3 galaxies: 

HDF$\_$PM3$\_$5 (J123635.6$+$621424, $z=$ 2.010), 

HDF$\_$PS3$\_$10 (J123642.1$+$621545, $z=$ 0.857),

HDF$\_$PM3$\_$21 (J123646.4$+$621529, $z=$ 0.851). 

Galaxies detected in the ultra-hard X-ray band (4-8 keV) or with
$L_{\rm X}\geq$ 10$^{43}$ erg s$^{-1}$ were classified as
AGNs. Galaxies with $L_{\rm X}\leq$ 10$^{40}$ erg s$^{-1}$ or which
could be fitted with the SED of M 82 or Arp 220 were classified in the
starburst category.

Finally only (12$\pm$5)\,$\%$ (5/41 galaxies, poissonian error bar) of
the ISOCAM-HDFN galaxies are found to be powered by an active
nucleus. These AGNs contribute to (18$\pm$7)\,$\%$ of the sum of the
flux densities (the IGL) of all 41 ISOCAM-HDFN galaxies. Since the
active nuclei in these galaxies may not be responsible for the
totality of the IR emission, this result suggests that at most
(18$\pm$7)\,$\%$ of the 15\,$\mu$m IGL due to galaxies with $S_{15}$=
0.1-0.5 mJy is produced by AGN activity.

The $S_{15}$= 0.5--3 mJy flux density range is covered by the ISOCAM
image of the Lockman Hole, where Fadda et al. (2001) cross-correlated
ISOCAM with the XMM-Newton X-ray observatory catalog of sources
brighter than $S$[2-10 keV]$\simeq$ 1.4$\times$10$^{-15}$ erg
cm$^{-2}$ s$^{-1}$ and $S$[5-10 keV]$\simeq$ 2.4$\times$10$^{-15}$ erg
cm$^{-2}$ s$^{-1}$ (Hasinger et al. 2001).  Among 103 ISOCAM sources
with a flux density ranging from 0.5 to 3 mJy, 13 sources were found
to be AGN dominated by Fadda et al. (2001), i.e. (13$\pm$4)\,$\%$.
Such galaxies could be fitted by the SEDs of either type-I or type-II
AGNs by Franceschini et al. (2002).  These AGNs contribute to
(15$\pm$5)\,$\%$ of the 15\,$\mu$m IGL due to galaxies in the 0.5--3
mJy flux range. 

The combination of the HDFN and Lockman Hole cover a flux density
range of 0.1-3 mJy where 70\,$\%$ of the 15\,$\mu$m IGL is measured.
In this range, these two fields suggest that AGNs contribute to at
most (17$\pm$6)\,$\%$. It must be noted that the XMM-Newton and
Chandra experiments were limited to energies below 10 keV. Some harder
X-ray AGNs may not have been detected by these instruments yet and may
be common with our ISOCAM source list. The presence of such objects
will remain will highly uncertain until the launch of the next
generation X-ray experiments (XEUS, Constellation X). We will assume
in the following that the AGN fraction computed in the HDFN (which is
larger than in the Lockman Hole) applyies for the 15\,$\mu$m IGL as a
whole.

\begin{table*}
\caption[]{Properties of the ISOCAM galaxies detected above 0.1 mJy in
the HDFN+FF (40 galaxies with spectroscopic redshift within a total of
41 galaxies). Col.(1): the galaxies have been divided into 5
categories (see text). Col.(2): median redshift. Col.(3): logarithm of
the median 15\,$\mu$m luminosity ($L_{15}=~\nu
L_{\nu}$[15\,$\mu$m]). Col.(4): logarithm of the median IR luminosity
(8-1000\,$\mu$m), calculated from the multi-template technique.
Col.(5): median star formation rate, calculated from $L_{\rm IR}$
using Eq.~(\ref{eqSFR}). The 1-$\sigma$ error bars in Cols.(3),(4),(5)
are the median of the 1-$\sigma$ error bars for each individual
galaxy.  They characterize the uncertainty on $L_{15}$, $L_{\rm IR}$
and $\left< SFR \right>$ for individual galaxies, not their dispersion
around the median value. Col.(6): number of galaxies in each
category. Col.(7): relative contribution of each galaxy type to the
15\,$\mu$m IGL ($IGL_{15}$) produced by the sum of all 40
galaxies. The mean and median values given in the last two lines
concern only the 35 star forming galaxies.}
\label{TAB:hdfn}
\begin{tabular}{|l|lrrcrr|}
\hline 
Type    &$\bar{ z}$& $log(L_{15})$& $log(L_{\rm IR})$ &$\left< SFR \right>$  &Nb& $\%\,IGL_{15}$ \\
        &   &($L_{\sun}$)  & ($L_{\sun}$)  &($M_{\sun}/yr$)       &  &           \\
\hline
ULIGs      & 1.2 & 11.3$\pm$ 0.1 & 12.3$\pm$ 0.2 & 332$\pm$148 &  6 & 17 \\
LIGs       & 0.8 & 10.5$\pm$ 0.1 & 11.5$\pm$ 0.2 &  56$\pm$26  & 20 & 44 \\
Starbursts & 0.5 &  9.8$\pm$ 0.1 & 10.7$\pm$ 0.2 &   9$\pm$4   &  5 &  9 \\
Normal     & 0.1 &  8.3$\pm$ 0.1 &  9.4$\pm$ 0.2 & 0.4$\pm$0.2 &  4 & 11 \\
AGNs       & 0.9 & 11.1$\pm$ 0.7 & 11.5$\pm$ 0.7 &      -      &  5 & 19 \\
\hline
Median     & 0.8 & 10.4$\pm$ 0.1 & 11.4$\pm$ 0.2 &  41$\pm$19  & 35 & 100 \\
Mean       & 0.8 & 10.8$\pm$ 0.1 & 11.8$\pm$ 0.2 & 109$\pm$51  & 35 & 100 \\
\hline
\end{tabular}
\end{table*}
\subsection{Rest-frame 15\,$\mu$m luminosity of ISOCAM galaxies}
\label{kcor}
In this section, we compare three different techniques to estimate the
rest-frame 15\,$\mu$m luminosity of ISOCAM galaxies. The first and
simplest technique consists in using a single template SED to compute
the k-correction. We chose the SED of M 82
(Fig.~\ref{FIG:sedsm82a220}a) which falls on the correlations plotted
in the Fig.~\ref{FIG:correl} (crosses in Fig.~\ref{FIG:kcor}b). As in
Sect.~\ref{resolve}, we decided not to use the SED of Arp 220
(Fig.~\ref{FIG:sedsm82a220}a) because it is well outside the
correlations found for galaxies of similar luminosities with its large
FIR over MIR luminosity ratio.

In a second technique, that we will call ``closest filter'', we use
directly the correlations described in Sect.~\ref{MIRtracer}
(Eq.~(\ref{eq:7mic}) and Eq.~(\ref{eq:12mic})) between the rest-frame
15\,$\mu$m and 6.75 or 12\,$\mu$m luminosities. We first calculate the
rest-frame wavelength $\lambda_{\rm rf}$=15/(1+$z$)\,$\mu$m and find
which of the three filters ISOCAM-LW3, ISOCAM-LW2 and IRAS-12 has its
central wavelength the closest to $\lambda_{\rm rf}$ (see
Fig.~\ref{FIG:kcor}a). Then we apply the formula given in the
Sect.~\ref{form_correl} for this ``closest filter'' to compute the
rest-frame 15\,$\mu$m luminosity (filled dots in
Fig.~\ref{FIG:kcor}b). Since most ISOCAM-HDFN galaxies are located
around the median redshift $\bar{ z} \sim$ 0.8, the correlation
between LW3 and LW2 (Fig.~\ref{FIG:correl}d, Eq. (\ref{eq:7mic}) is
most often used.

The third technique is the multi-template technique already used in
Sect.~\ref{sensrad}, where we use the library of 105 template SEDs
from Chary \& Elbaz (2001). For a given galaxy of flux density
$S_{15}$ located at a given redshift $z$, we have computed
$S\arcmin_{15}$ for each one of the 105 template SED if it were
located at the same redshift $z$. The template for which the computed
$S\arcmin_{15}$ was the closest to the observed $S_{15}$ was then used
to compute the rest-frame 15\,$\mu$m luminosity, $\nu
L_{\nu}$[15\,$\mu$m] (open circles in Fig.~\ref{FIG:kcor}b), after a
normalization by a factor $S_{15}$/$S\arcmin_{15}$.

The rest-frame 15\,$\mu$m luminosities estimated using the three
techniques for the ISOCAM-HDFN galaxies are consistent within 20\,$\%$
(1-$\sigma$, see Fig.\ref{FIG:kcor}b). Above $z\sim 0.6$, the first
technique, which makes use of the SED of M 82, systematically
underestimates $\nu L_{\nu}$[15\,$\mu$m], as compared to the other two
techniques. This was to be expected since M 82 is a moderately
luminous IR galaxy with a lower FIR over 6.75\,$\mu$m luminosity ratio
than more luminous galaxies (see Fig.~\ref{FIG:correl}d).

In the following, we will use the multi-template technique to estimate
the IR luminosity and SFR of galaxies, because it is consistent with
the correlations between MIR and FIR luminosities and provides a
continuous interpolation between the 6.75, 12,15\,$\mu$m and FIR
luminosities based on observed spectra. Since the templates fit the
correlation between $L_{\rm IR}$ and $\nu L_{\nu}$[15\,$\mu$m], it is
equivalent to compute $\nu L_{\nu}$[15\,$\mu$m] first and then use
Eq.~(\ref{eq:15mic}) which links $L_{\rm IR}$ and $\nu
L_{\nu}$[15\,$\mu$m] or to estimate directly $L_{\rm IR}$ with the
best template SED.

\subsection{Luminosities and star formation rates}
\label{lumir}
\begin{figure}
\resizebox{\hsize}{!}{\includegraphics{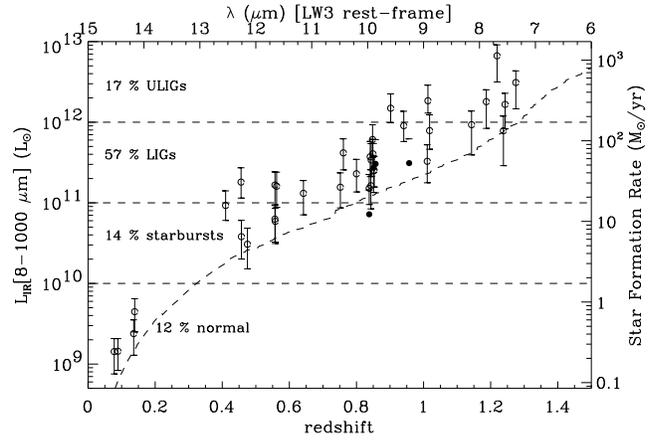}}
\caption{$L_{\rm IR}$[8-1000\,$\mu$m] and star formation rate (SFR)
versus redshift and $\lambda_{\rm rf}$ for the ISOCAM-HDFN galaxies
($S_{15}\geq$ 0.1 mJy). The 15\,$\mu$m completeness limit of 0.1 mJy
is materialized with a dashed line. Only the left axis is meaningful
for the five AGNs (filled dots, Sect.~\ref{agnfrac}). HDF\_PS3\_6a (no
spectroscopic redshift) is not included.}
\label{FIG:sfr}
\end{figure}
In this section, we use the multi-template technique to directly
estimate the $L_{\rm IR}$ and SFR of the ISOCAM-HDFN galaxies. The
results are shown in the Fig.~\ref{FIG:sfr} and
Table~\ref{TAB:hdfn}. The 1-$\sigma$ error bars in Fig.~\ref{FIG:sfr}
were computed from the dispersion measured in the correlations of the
Fig.~\ref{FIG:correl}. For example, for a galaxy located at $z\simeq$
1, since $\lambda_{\rm rf}$ is close to 7\,$\mu$m, the error bar on
$L_{\rm IR}$ is the quadratic sum of the error bar on
Eq.~(\ref{eq:7mic}), which gives $L_{15}$ from $L_{7}$, and of the
error bar on Eq.~(\ref{eq:15mic}), which gives $L_{\rm IR}$ from
$L_{15}$.

\begin{figure*}
\resizebox{\hsize}{!}{\includegraphics{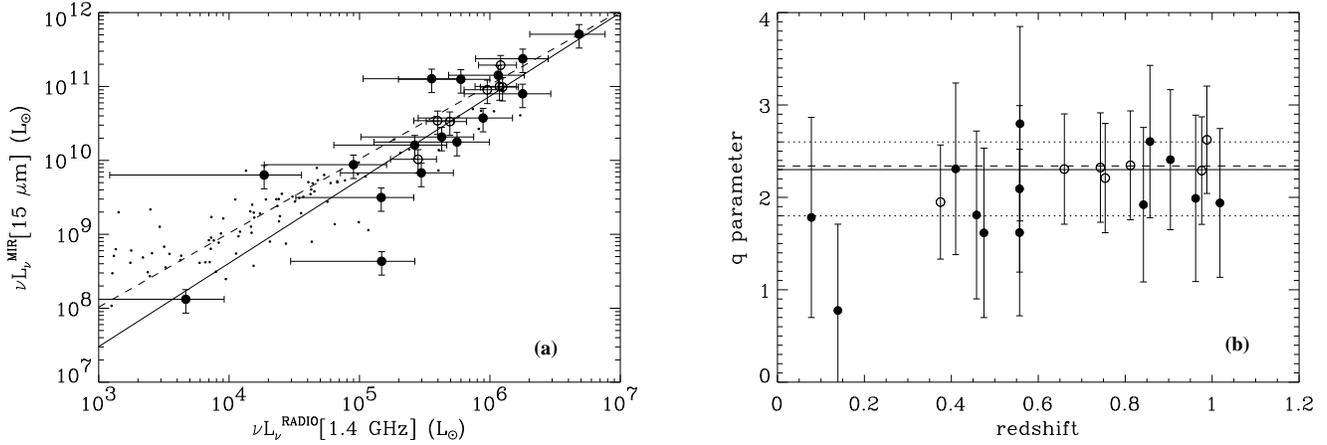}}
\caption[]{{\bf a)} 15\,$\mu$m versus radio continuum (1.4 GHz)
luminosities. Small filled dots: sample of 109 local galaxies from
ISOCAM and NVSS. Filled dots with error bars: 17 HDFN galaxies in
common between ISOCAM and VLA or WSRT (see text). Open dots with error
bars: 7 CFRS-14 galaxies in common between ISOCAM and VLA (Flores et
al. 1999). {\bf b)} ``$q$'' parameter as a function of redshift for
the HDFN and CFRS-14 galaxies. Plain line: median value of ``q''
(2.3$^{+0.3}_{-0.5}$). Dashed line: measured value from local galaxies
($q$= 2.34$\pm$0.01; Yun, Reddy \& Condon 2001). Dotted line:
1-$\sigma$ error bar on ``q'' for the 24 galaxies.}
\label{FIG:radio}
\end{figure*}
The dashed line on Fig.~\ref{FIG:sfr} corresponds to the sensitivity
limit of $S_{15}$= 0.1 mJy converted into a $L_{\rm IR}(z)$ as in the
Fig.~\ref{FIG:sensM82}. Table~\ref{TAB:hdfn} summarizes the properties
of the ISOCAM galaxies in each of these four galaxy categories plus
the AGN category.

About 75\,$\%$ of the galaxies dominated by star formation are either
LIGs or ULIGs, while the remaining $\sim$ 25\,$\%$ are nearly equally
distributed among either ``starbursts'' ($10^{11}> (L_{\rm
IR}/L_{\sun})\geq 10^{10}$) or ``normal'' galaxies ($L_{\rm IR}<$
10$^{10}~L_{\sun}$). The contribution of each galaxy type to the
15\,$\mu$m IGL ($IGL_{15}$) is given in the last column of the
Table~\ref{TAB:hdfn}. Luminous IR galaxies are responsible for about
60\,$\%$ of $IGL_{15}$ while AGNs contribute to about 20\,$\%$, the
remaining 20\,$\%$ being due to the normal and starburst galaxy
types. This repartition will be used in Sect.~\ref{cirb} to compute
the 140\,$\mu$m IGL ($IGL_{140}$) from ISOCAM galaxies.

The mean IR luminosity of the ISOCAM-HDFN galaxies is $L_{\rm
IR}\simeq$ 6$\times$10$^{11}$ $L_{\sun}$. It corresponds to a
$SFR\simeq$ 100 M$_{\sun}$ yr$^{-1}$ (from Eq.~(\ref{eqSFR})). In the
following section, we evaluate the robustness of this result with an
independent estimator of the IR luminosity: the radio luminosity.
\subsection{The radio-MIR correlation}
\label{radio}
The radio continuum is also a tracer of the bolometric IR luminosity
of star forming galaxies as a result of the tight correlation between
IR and 1.4 GHz radio continuum in local galaxies (Condon 1992,
Yun, Reddy \& Condon 2001; see Sect.~\ref{sensrad}). The origin of
this correlation remains unclear, but it is generally assumed that
massive stars are both responsible for the UV photons that heat dust
before it radiates in the IR and for the synchrotron acceleration of
electrons, producing the radio continuum, when they explode as
supernovae.

In Fig.~\ref{FIG:radio}a, we have plotted the 1.4 GHz and 15\,$\mu$m
luminosities of the 109 local galaxies (small filled dots) from the
ISOCAM sample described in Sect.~\ref{data}, which were detected in
the NRAO VLA Sky Survey (NVSS, Condon et al. 1998). As expected both
luminosities are correlated with each other since both the 1.4 GHz and
15\,$\mu$m luminosities are correlated with $L_{\rm IR}$.

Half of the 35 star forming HDFN-ISOCAM galaxies with spectroscopic
redshift (AGNs excluded) present flux densities larger than
40\,$\mu$Jy at 1.4 GHz (5-$\sigma$) in the VLA and WSRT catalogs from
Richards (2000) and Garrett et al. (2000) respectively. Their
rest-frame 15\,$\mu$m luminosities were computed in the
Sect.~\ref{kcor}. Their rest-frame 1.4 GHz luminosities were computed
assuming a power-law as in Sect.~\ref{sensrad}: $S_{\nu}\propto
\nu^{-\alpha}$, where $\alpha$= 0.8$\pm$0.15 as suggested for star
forming galaxies in Yun, Reddy \& Condon (2001). The rest-frame
15\,$\mu$m and 1.4 GHz luminosities of these 17 HDFN-ISOCAM galaxies
are plotted as filled dots with error bars in the
Fig.~\ref{FIG:radio}a. The error bars on $\nu L_{\nu}$[1.4 GHz] were
computed from the quadratic sum of the error bars on the measurement
of the radio flux densities plus the error bar on $\alpha$. Nine
galaxies common in both catalogs from Richards (2000) and Garrett et
al. (2000) present up to 40\,$\%$ differences in their radio flux
densities.  For these galaxies, we used the mean value and included
the difference between both measurements in the error bars.

We have also included 7 galaxies from the CFRS-14 field (Flores et
al. 1999), which rest-frame luminosities were computed with the same
techniques (open dots).

The MIR and radio luminosities of this sample of 24 distant dusty
galaxies (${\bar z}\simeq$ 0.8) are also strongly correlated with each
other.  This is the first time that such correlation is found at these
redshifts.  The plain line in the Fig.~\ref{FIG:radio}a is a power-law
fit to this correlation. For comparison, we have plotted in dashed
line the correlation that one would expect from the combination of
Eq.~(\ref{eq:15mic}) ($L_{\rm IR}$ from $L_{15}$) and the radio
``$q$'' parameter (see Sect.~\ref{sensrad}). The slope of the
correlation observed for distant galaxies is marginally steeper than
this dashed line. It is also slightly steeper than the correlation
found for local galaxies (small filled dots), but the difference is
too marginal with respect to the number of galaxies at high
luminosities for detailed interpretation.

In Fig.~\ref{FIG:radio}b, the ``$q$'' parameter for the 24 ISOCAM-HDFN
and CFRS-14 galaxies is plotted as a function of redshift. The ``$q$''
parameter is the logarithm of the ratio of $L_{\rm FIR}$ over $L$(1.4
GHz) (see Eq.~(\ref{eq:qpar})). $L_{\rm IR}$ was estimated as described
in Sect.~\ref{lumir}, then converted into $L_{\rm FIR}$ from
Eq.~(\ref{eq:irfir}). The median value of the ``$q$'' parameter for
the 24 galaxies is: $q$= 2.3$^{+0.3}_{-0.5}$ (plain and dotted lines
in Fig.~\ref{FIG:radio}b), in perfect agreement with the local value
of $q$= 2.34$\pm$0.01 (Yun, Reddy \& Condon 2001, dashed line in
Fig.~\ref{FIG:radio}b).

This study shows that the IR luminosities estimated from the ISOCAM
15\,$\mu$m flux densities using the MIR-FIR correlations are perfectly
consistent with those estimated from the radio. Although it is not
clear whether the radio versus IR correlation also applyies up to
$z\sim$ 1, this result independently validates our estimate of the
bolometric IR luminosity of the ISOCAM galaxies.

\subsection{Evolution of the infrared volume emissivity with redshift}
\label{ldens}
\begin{figure*}
\resizebox{\hsize}{!}{\includegraphics{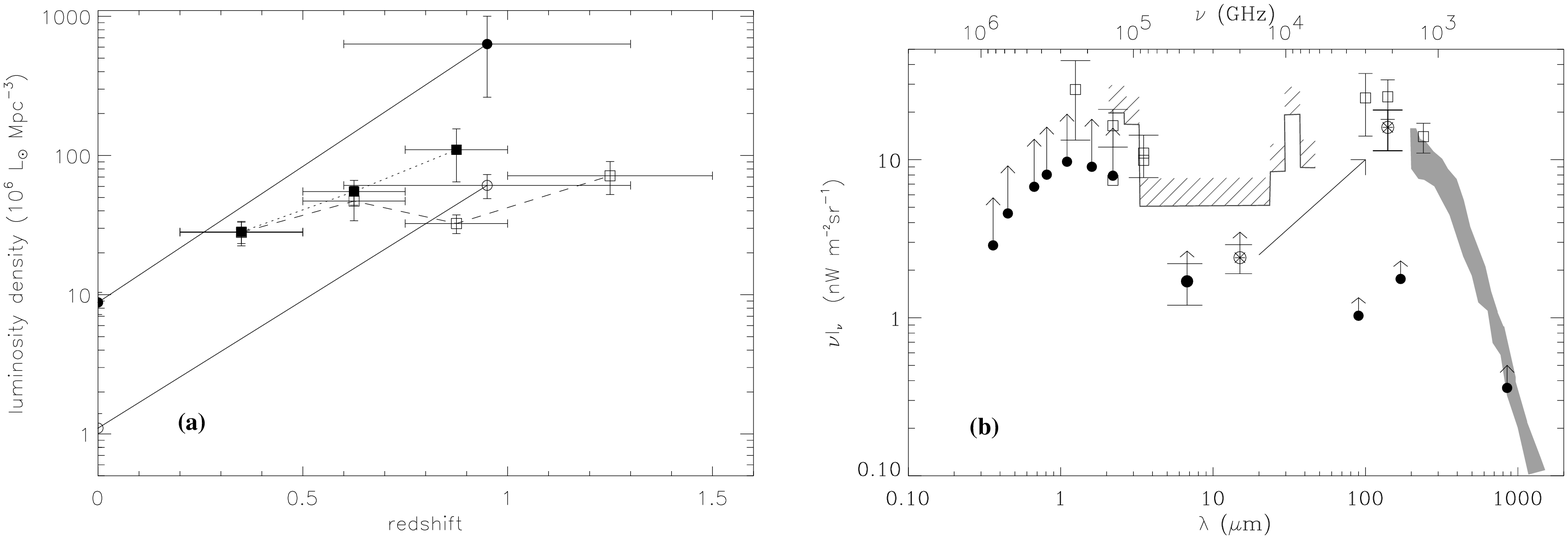}}
\caption{{\bf a)} FIR (filled circles), MIR-15\,$\mu$m (open
circles) and UV-2800$\AA$ luminosity density (in $L_{\sun}$ Mpc$^{-3}$)
as a function of redshift. UV-2800$\AA$: open squares from Cowie,
Songaila \& Barger (1999); filled squares from Lilly et
al. (1996). {\bf b)} Integrated Galaxy Light (IGL, filled dots) and
Extragalactic Background Light (EBL, open squares, grey area) from the
UV to sub-millimeter (see Table~\ref{TAB:ebl} for the origin of the
data). EBL measurements from COBE: 200-1500\,$\mu$m EBL from
COBE-FIRAS (grey area, Lagache et al. 1999), 1.25, 2.2, 3.5, 100,
140\,$\mu$m EBL from COBE-DIRBE (open squares, see
Table.~\ref{TAB:ebl}). IGL in the U,B,V,I,J,H,K bands from Madau \&
Pozzetti (2000). The upper end of the arrows indicate the revised
values suggested by Bernstein et al. (2001, factor two higher). Our
estimate of the 15\,$\mu$m IGL (2.4 $\pm$ 0.5 nW m$^{-2}$ sr$^{-1}$)
is marked with a surrounded star. 6.75\,$\mu$m (ISOCAM-LW2 filter) IGL
from Altieri et al. (1999, filled dot). Hatched upper limit from Mkn
501 (Stanev \& Franceschini 1998). The arrow from 15 to 140\,$\mu$m
indicates our computation of the 140\,$\mu$m IGL due to ISOCAM
galaxies.}
\label{FIG:ldensbkg}
\end{figure*}
We have seen that the excess of faint MIR sources in number counts was
due to the presence of distant ($z\sim$ 1) luminous IR galaxies.  As a
consequence, the amount of star formation per comoving volume hidden
by dust must have rapidly decreased from z$\sim$1 to 0. Indeed, the
ISOCAM-HDFN galaxies with 0.6 $\leq z \leq$ 1.3 and $L_{\rm IR}\geq$
10$^{11}~L_{\sun}$ (LIGs and ULIGs; AGNs excluded) produce a
15\,$\mu$m luminosity density of ${\cal L}_{15}\left({\bar z} \sim
1\right)\simeq (60 \pm 12)\times10^6~L_{\sun}$ Mpc$^{-3}$, while in
the local universe luminous IR galaxies only make ${\cal L}_{15}(z
\sim 0)\simeq (11 \pm 3)\times10^5~L_{\sun}$ Mpc$^{-3}$ (see
Table~\ref{TAB:comov}). The comoving luminosity density produced by
luminous IR galaxies at 15\,$\mu$m was therefore about 55 times larger
at $z\sim$ 1 than in the local universe. The 15\,$\mu$m luminosity
densities at $z\sim$ 0 were calculated using the local luminosity
function (LLF) from Xu et al. (1998). In order to estimate the
contribution of luminous IR galaxies, we used the Eq.(\ref{eq:15mic})
to convert the IR luminosity cut-off of $L_{\rm IR}=~10^{11}~L_{\sun}$
into a 15\,$\mu$m luminosity cut-off of $L_{15}\sim~10^{10}~L_{\sun}$.
The 15\,$\mu$m luminosity density that we have computed is consistent
with the one measured at 12\,$\mu$m from IRAS, ${\cal
L}_{12}^{z\sim0}\sim (1.7 \pm 0.5)~\times10^6~L_{\sun}$ Mpc$^{-3}$,
converted to 15\,$\mu$m using Eq.(\ref{eq:12mic}).

If we now consider the bolometric IR luminosity (from 8 to
1000\,$\mu$m) of ISOCAM galaxies, as estimated in the
Sect.~\ref{bolIR}, we find that the comoving density of IR luminosity
radiated by dusty starbursts was about (70 $\pm$ 35) times larger at
${\bar z}\sim$ 1 than today (computed from the LLF of Soifer et
al. 1987). Since the IR luminosity is directly proportional to the
extincted star formation rate of a galaxy, this means that the
comoving density of star formation taking place in luminous IR
galaxies was about (70 $\pm$ 35) times larger at $z\sim$ 1 than
today. In case of a pure density evolution proportional to
(1+$z$)$^n$, this would translate into a value of $n\simeq$
6. However, we want to emphasize that we are only considering here the
galaxies detected by ISOCAM, not the full luminosity function. For
comparison, the $B$-band (0.44\,$\mu$m) luminosity density was only
about three times larger at $z\sim$1 than today.

The redshift evolution of the comoving density of IR luminosity is
compared to the UV (2800\AA) one in Fig.~\ref{FIG:ldensbkg}a. Both
wavelengths exhibit similar luminosity densities at low redshift but
the IR rises faster than the UV and reaches a larger value at $z\sim$
1, implying that a much larger fraction of star formation was hidden
by dust at $z\sim$ 1 than today.

Finally, we note that the projected density of galaxies detected in
the $B$-band in the HDFN (529 galaxies/$\arcmin^2$ with $B_{\rm AB}\leq$
29, Pozzetti et al. 1998) is 330 times greater than the projected
density of ISOCAM galaxies (1.6 sources/$\arcmin^2$, with $S_{15}\geq$
0.1 mJy or $AB$(15\,$\mu$m)$\leq$ 18.9). However the ISOCAM galaxies
produce a 15\,$\mu$m IGL which is only twice lower than the $B$-band
IGL ($IGL_B\sim~4.57^{+0.73}_{-0.47}$ nW m$^{-2}$ sr$^{-1}$, Madau \&
Pozzetti 2000). This confirms that the very luminous galaxies detected
in the MIR radiate mostly in the IR.

\begin{table}
\begin{tabular}{|l|ccccc|}
\hline
	   & \multicolumn{2}{c}{15\,$\mu$m} &  \multicolumn{2}{c}{8-1000\,$\mu$m} & 0.44\,$\mu$m \\
           & All         &     LIGs         &  All             &     LIGs         & All          \\
\hline
$z\sim$0   & 7.5$\pm$1.6 &  1.1$\pm$0.3     &   120$\pm$30     &    8.8$\pm$1.6   &  82$\pm$2    \\
$z\sim$1 &     -         &   61$\pm$12      &       -          &    632$\pm$291   & 240$\pm$100  \\
\hline
\end{tabular}
\caption{Comoving luminosity density in units of
$10^6~L_{\sun}~Mpc^{-3}$ in the local universe and at $z\sim$ 1 (0.6
$\leq\,z\geq$ 1.3).  Columns ``All'': contribution of galaxies of all
luminosities are considered. Columns ``LIGs'': only galaxies with
$L_{\rm IR}\geq~10^{11}~L_{\sun}$. AGNs were excluded for the
computation of the IR values. The $B$-band (0.44\,$\mu$m) values are
from Lilly et al. (1996) at $z\sim$0 and Connolly et al. (1997) at
$z\sim$ 1 (mean of the 0.5--1 and 1--1.5 redshift bins).}
\label{TAB:comov}
\end{table}

\section{Contribution of ISOCAM galaxies to the cosmic infrared background} 
\label{cirb}
Fig.~\ref{FIG:ldensbkg}b summarizes the current estimates of the
cosmic background for $\lambda=$ 0.1 to 1500\,$\mu$m. Lower limits to
the EBL (filled dots) come from integrating galaxy fluxes (IGL), while
direct measurements of the EBL come from the COBE experiments DIRBE
(open squares) and FIRAS (grey area, $\lambda>$ 200\,$\mu$m, Lagache
et al. 1999). Table~\ref{TAB:ebl} gives the origin of the data used in
Fig.\ref{FIG:ldensbkg}b. Three independent studies obtain the same FIR
to sub-millimeter EBL above $\lambda\sim 140\,\mu$m. We did not
consider here the EBL at $\lambda=\,60\,\mu$m (Finkbeiner, Davis \&
Schlegel 2000) because it strongly depends on the remains controversial
(see Puget \& Lagache 2001).

\begin{table}
\begin{tabular}{|l|l|l|l|l|l|l|}
\hline
Filter & $\lambda$ & $\nu I_{\nu}^{(*)}$    & $[\sigma^+,\sigma^-]^{(a)}$ & $S_{min}$& AB & Ref \\
       & $\mu$m   &                  &            & $mJy$ & mag &      \\
\hline
U      & 0.36      & 2.87           & [0.58,0.42]  & 0.023 & 28    & 1  \\
B      & 0.45      & 4.57           & [0.73,0.47]  & 0.009 & 29    & 1  \\
V      & 0.67      & 6.74           & [1.25,0.94]  & 0.002 & 30.5  & 1  \\
I      & 0.81      & 8.04           & [1.62,0.92]  & 0.009 & 29    & 1  \\
J      & 1.1       & 9.71           & [3.00,1.90]  & 0.009 & 29    & 1  \\
H      & 1.6       & 9.02           & [2.62,1.68]  & 0.009 & 29    & 1  \\
K      & 2.2       & 7.92           & [2.04,1.21]  & 0.23  & 25.5  & 1  \\
LW2    & 6.75      & 1.7            & 0.5          & 0.03  & 20.2  & 2  \\
{\bf LW3}&{\bf 15} &{\bf 2.4} &{\bf0.5}&{\bf 0.05}&{\bf 19.7}&    \\
PHOT   &  90       & 1.03           &              & 1.5e3 & 11    & 3 \\
PHOT   & 170       & 0.88           &              & 1.5e3 & 11    & 3 \\
PHOT   & 170       & 1.76           &              & 1.2e3 & 11    & 4 \\
SCUBA  & 850       & 0.36           &              & 2     & 15.6  & 5 \\
SCUBA$^{(l)}$& 850     & 0.5            & 0.2          & 0.5   & 17.2  & 6  \\
\hline
\hline
       & 1.25      & 27.8           & 14.5         &       &       & 7 \\
       & 1.25      & 22.9           &  7.0         &       &       & 8  \\
       & 2.2       & 20.4           &  4.9         &       &       & 8  \\
K      & 2.2       & 7.4            & 6.9          &       &       & 9 \\
K      & 2.2       & 16.4           & 4.4          &       &       &10 \\
       & 3.5       & 7.4            & 6.9          &       &       & 9 \\
       & 3.5       & 11.0           & 3.3          &       &       & 8 \\
DIRBE  & 60        & 28.1           & 8.8$^{(b)}$  &       &       &11 \\
DIRBE  & 100       & 24.6           & 10.5$^{(c)}$ &       &       &11 \\
DIRBE  & 100       & 23.4           & 6.3          &       &       &12 \\
DIRBE  & 140       & 25.0           & 7.0          &       &       &11 \\
DIRBE  & 140       & 24.2           & 11.6         &       &       &12 \\
DIRBE  & 240       & 14.0           & 3.0          &       &       &11 \\
DIRBE  & 240       & 11.0           & 6.9          &       &       &12 \\
FIRAS  & 850       & 0.5            & 0.2          &       &       &13 \\
\hline
\end{tabular}
\caption{Integrated Galaxy Light (IGL, upper half of the table) and
Extragalactic Background Light (EBL, lower half of the table), $\nu
I_{\nu}$, in units of $(*)$ nW m$^{-2}$ sr$^{-1}$. $(a)$ 1-$\sigma$
error bars. $(b)$,$(c)$ include the systematic error quoted by the
authors. Ref: (1) Madau \& Pozzetti (2000), (2) Altieri et al. (1999),
(3) Matsuhara et al. (2000), (4) Puget et al. (1999), (5) Barger,
Cowie \& Sanders (1999), (6) Blain et al. (1999b), lensed galaxies
$^(l)$, (7) Wright (2000), (8) Cambr\'esy et al. (2001), (9) Dwek \&
Arendt (1998), (10) Gorjian, Wright \& Chary (2000), see also Wright
\& Reese (2000), Wright (2000), (11) Finkbeiner, Davis \& Schlegel
(2000) and Hauser et al. (1998) for the 140 and 240\,$\mu$m values,
(12) Lagache et al. (2000), (13) Lagache et al. (1999), Fixsen et
al. (1998)}
\label{TAB:ebl}
\end{table}

Below $\lambda\sim~5\,\mu$m, the cosmic background originates from
direct stellar light, while above this wavelength, it comes from
either stellar or AGN light reprocessed by dust. In the dust domain,
i.e. from $\lambda\sim$ 5 to 1500\,$\mu$m, the cosmic background peaks
around $\lambda\sim~140\,\mu$m. It is not directly measured over this
whole wavelength range, but below $\lambda\sim~40\,\mu$m and down to
about 2\,$\mu$m, an upper limit is set by the observations of the TeV
outburst of Mkn 501 (see Sect.~\ref{ebl15}). 

The 15\,$\mu$m IGL (IGL$_{15}$ = (2.4$\pm$0.5) nW m$^{-2}$ sr$^{-1}$,
see Sect.~\ref{ebl15}) is marked with a surrounded star on the
Fig.~\ref{FIG:ldensbkg}b. It is about ten times lower than the peak of
the CIRB at $\lambda_{\rm max}\sim~140\,\mu$m. At the median redshift
of the ISOCAM galaxies, $\bar{z}\sim$ 0.8 (Sect.~\ref{nature}), the
rest-frame wavelengths corresponding to the observed 12-18\,$\mu$m
(ISOCAM-LW3 bandpass) and 140\,$\mu$m wavelengths, are 6.5-9.8\,$\mu$m
and $\sim$\,80\,$\mu$m. These rest-frame wavelengths are close to the
ISOCAM-LW2 (5-8.5\,$\mu$m) and IRAS (60, 100\,$\mu$m) bands. We have
seen in the Fig.~\ref{FIG:correl}d that the luminosities of local
galaxies in both bands correlate with each other. If this correlation
remains valid up to $z\sim$ 1, then it implies that we can compute the
140\,$\mu$m IGL due to ISOCAM galaxies. There is no obvious reason why
the SEDs of galaxies at $z\sim$ 1 would exhibit very different IR
SEDs. For example, the SED of the extremely red object HR 10 ($z\sim$
1.44) is very similar to the one of Arp 220 normalized by a factor of
$\sim$ 4 (Elbaz et al. 2002).

We propose here to use the multi-template technique, as in
Sect.~\ref{bolIR}, to compute the contribution of ISOCAM galaxies to
the 140\,$\mu$m background. We will first separate the relative
contributions to IGL$_{15}$ from ULIGs, LIGs, SBs, normal galaxies and
AGNs as estimated in the Sect.~\ref{bolIR}. For each galaxy type, a
median redshift, $z_{med}$, and the ratio ${\cal R}= \nu
S_{\nu}[140/(1+z_{med})]$/$\nu S_{\nu}[15/(1+z_{med})]$ are estimated
based on the HDFN sample (see Table~\ref{TAB:frac}). Although a larger
sample would improve the statistical reliability of this computation,
this choice is justified by the fact that the bulk of the 15\,$\mu$m
IGL is produced by galaxies with flux densities in the range
$S_{15}\simeq$ 0.35-1 mJy as the HDFN and CFRS-14 galaxies, which
share similar redshifts distributions. It is important to separate the
contribution from galaxies of different luminosities since the IR
versus 6.75\,$\mu$m luminosity correlation presents a slope larger
than one (Fig.~\ref{FIG:correl}d). The contribution from AGNs is
computed assuming that they all share the SED of NGC 1068. This is a
conservative hypothesis since NGC 1068 exhibits the lowest FIR over
MIR ratio that we know.

The results are summarized in the Table~\ref{TAB:cirb}. We first
separate the contribution of each type of galaxy to the 15\,$\mu$m IGL
(Col.(3),(4)). The 140\,$\mu$m IGL is then computed by converting each
contribution to the 15\,$\mu$m IGL into a contribution at 140\,$\mu$m,
by multiplying by ${\cal R}$ each time. We find a total of
$IGL_{140}\simeq$ (16$\pm$5) nW m$^{-2}$ sr$^{-1}$ (see
Fig.~\ref{FIG:ldensbkg}b), which corresponds to (65$\pm$35)\,$\%$ of
the observed value from COBE-DIRBE ($EBL_{140}$= (25$\pm$7) nW
m$^{-2}$ sr$^{-1}$, Hauser et al. 1998, Lagache et al. 2000,
Finkbeiner et al. 2000).

Hence the galaxies detected in ISOCAM deep surveys are found to be
responsible for the bulk of the CIRB. About half of this 140\,$\mu$m
IGL is due to LIGs and about one third from ULIGs. The contribution of
AGNs is estimated to be as low as $\sim$ 4\,$\%$ of the 140\,$\mu$m
IGL. It could be as high as $\sim$ 20\,$\%$ if instead of NGC 1068, we
had used the SED of NGC 6240. In this case, ISOCAM galaxies would
produce a $IGL_{140}\simeq$ (19$\pm$5) nW m$^{-2}$ sr$^{-1}$.  If we
had included the contribution from galaxies with flux densities
between 30 and 50\,$\mu$Jy only detected through lensing magnification
(see Sect.~\ref{ebl15}), $IGL_{140}$ would be larger by an extra
$\sim$ 4 nW m$^{-2}$ sr$^{-1}$.

At the mean redshift of $z\simeq$ 0.6 where the bulk of the $B$-band
IGL is produced (Pozzetti \& Madau 2001), light was radiated in the UV
in order to be observed at 0.44\,$\mu$m. As we have seen, the bulk of
the 140\,$\mu$m is also produced in the same redshift range. Hence the
140\,$\mu$m and $B$-band probe the UV emission from stars with and
without dust extinction. The ratio of $IGL_{140}$ over $IGL_{B}$
should therefore give a rough approximation of the ratio of extincted
over non extincted star formation around $z\sim$ 0.5-1. This ratio is
close to 5, which suggests again that the bulk of the UV photons
radiated by young stars in this redshift range was strongly affected
by dust extinction. This value is consistent with the one obtained
by Chary \& Elbaz (2001) or Franceschini et al. (2001).

\begin{table}
\begin{center}
\caption{Contribution of the ISOCAM galaxies to the 140\,$\mu$m
extragalactic background light. Col.(1) galaxy type: AGNs and star
forming galaxies above and below $log(L_{\rm IR}/L_{\sun})=$ 10, 11
and 12 (see Sect.~\ref{lumir}). $\bar{z}$, in Col.(2), is the median
redshift for each galaxy type. ``$\%$15'', in Col.(3), is the
fractional contribution of each galaxy type to $IGL_{15}$
(Eq.(\ref{eq:igl})). Col.(4), $IGL_{15}$ for each galaxy type in nW
m$^{-2}$ sr$^{-1}$. Col.(5), ${\cal R}=\,\nu S_{\nu}[140\,\mu$m]$/\nu
S_{\nu}[15\,\mu$m]. For AGNs, we used the SED of NGC
1068. IGL$_{140}$, in Col.(4), is the 140\,$\mu$m IGL produced by each
galaxy type in nW m$^{-2}$ sr$^{-1}$ and $\%$140, in Col.(5), is the
fractional contribution to the 140\,$\mu$m IGL for each galaxy type.}
\begin{tabular}{|l|rrcrrr|}
\hline
       &         &              &             &          &             &                 \\
Type   &$\bar{z}$&$\%$15        &IGL$_{15}$   &${\cal R}$& IGL$_{140}$ & $\%$140         \\
       &         &              &             &          &             &                 \\
\hline 
ULIGs  &1.2      &  17          &0.4$\pm$0.1  &14$\pm$6  & 5.7$\pm$2.8 & 36 \\
LIGs   &0.8      &  44          &1.1$\pm$0.2  & 7$\pm$3  & 7.7$\pm$3.6 & 48 \\
SBs    &0.5      &   9          &0.2$\pm$0.1  & 4$\pm$1  & 1.0$\pm$0.4 &  6 \\
normal &0.1      &  11          &0.3$\pm$0.1  & 4$\pm$1  & 0.9$\pm$0.5 &  6 \\
AGNs   &1.0      &  19          &0.5$\pm$0.1  &1.4       & 0.7$\pm$0.1 &  4 \\
\hline 	           	           	       	          	        
Total  &0.8      & 100          &2.4$\pm$0.5  & 7$\pm$3  &16.0$\pm$4.6 &100 \\
\hline
\end{tabular}
\end{center}
\label{TAB:cirb}
\end{table}

\section{Conclusions}
\label{conclusions}
The cosmic IR background is a fossil record of the light radiated by
galaxies since their formation. It peaks around $\lambda_{\rm
max}\simeq$ 140\,$\mu$m whereas the spectral energy distribution of
galaxies peaks above 60\,$\mu$m. This suggests that the bulk of the
cosmic IR background is due to galaxies located below $z\sim$ 1.3. We
have shown that the best technique currently available to unveil dusty
galaxies up to $z\sim$ 1.3 is provided by the ISOCAM MIR extragalactic
surveys.

We have computed the contribution of ISOCAM galaxies to the 15\,$\mu$m
background, the 15\,$\mu$m integrated galaxy light, and found a value
of $IGL_{15}\simeq$ (2.4$\pm$0.5) nW m$^{-2}$ sr$^{-1}$. This is about
ten times below the cosmic background measured by COBE at
$\lambda_{\rm max}\simeq$ 140\,$\mu$m: $EBL_{140}\simeq$ (25$\pm$7) nW
m$^{-2}$ sr$^{-1}$.

We have demonstrated that the MIR luminosities at 6.75, 12 and
15\,$\mu$m were correlated with each other and with the bolometric IR
luminosity for local galaxies. This suggests that the contribution of
ISOCAM galaxies to the CIRB can be computed from $IGL_{15}$, unless
distant galaxies SEDs strongly differ from local ones. 

The redshift distribution of ISOCAM galaxies was measured from the
spectroscopically complete sample of galaxies in the region of the
HDFN. This redshift distribution is consistent with the twice larger
sample of ISOCAM galaxies detected in the CFRS fields CFRS-14 and
CFRS-03 (Flores et al. 1999, 2002). At the median redshift of ${\bar
z}\simeq$ 0.8, the observed 15 and 140\,$\mu$m wavelengths correspond
to about 7\,$\mu$m (ISOCAM-LW2 filter) and 80\,$\mu$m (IRAS bands) in
the rest-frame of the galaxies. Luminosities at both wavelengths are
correlated (see Fig.~\ref{FIG:correl}d). If the correlations between
MIR and FIR luminosities remain valid up to $z\sim$ 1, then they can
be used to compute $IGL_{140}$.  We have checked with a sample of
galaxies detected both in the MIR with ISOCAM and in the radio with
the VLA and WSRT, that the MIR-FIR and radio-FIR correlations are
consistent up to $z\sim$ 1.  This comparison independently validates
our estimate of the bolometric IR luminosity of the ISOCAM galaxies,
although it is not clear whether the radio-FIR correlation works also
up to $z\sim$ 1.

The fraction of active nuclei responsible for the 15\,$\mu$m
luminosity of ISOCAM galaxies was estimated from the deepest soft and
hard X-ray surveys available at present by the XMM-Newton and Chandra
X-ray observatories in the Lockman Hole and HDFN respectively (Fadda
et al. 2001). It was found that about (12$\pm$5)\,$\%$ of the ISOCAM
galaxies are powered by an AGN and that the AGN contribution to
$IGL_{15}$ was about (17$\pm$6)\,$\%$. The AGN contribution to
$IGL_{140}$ was found to be as low as $\sim$ 4\,$\%$ assuming that
they all share the SED of the local Seyfert 2, NGC 1068. This is a
conservative choice since NGC 1068 presents the flattest IR SED that
we know. However, we note that the cosmic X-ray background (CXB) peaks
around 30 keV (see Fig.1 of Wilman, Fabian \& Nulsen 2000), while both
XMM-Newton and Chandra were limited to energies below 10 keV. It is
therefore possible that a population of hard X-ray AGNs was missed by
these surveys. But as claimed by the authors of these deep X-ray
surveys, the bulk of the CXB had been resolved into individual
galaxies in the Lockman Hole and HDFN images. Moreover, using
estimates of the present comoving density of black holes, Madau \&
Pozzetti (2000) calculated that less than 20\,$\%$ of the CIRB could
be due to dusty AGNs.

For the remaining star forming galaxies, we used a library of template
SEDs, reproducing the MIR-FIR correlations, to compute their IR
luminosity and contribution to $IGL_{140}$. We find that LIGs and
ULIGs produce about 60\,$\%$ of $IGL_{15}$. The comoving density of IR
luminosity produced by these luminous IR galaxies was about
(70$\pm$35) times larger at $z\sim$ 1 than today, while in the same
redshift interval the $B$-band or UV luminosity densities only
decreased by a factor $\sim$ 3. Since the IR luminosity measures the
dusty star formation rate of a galaxy, this also implies that the
comoving density of star formation, due to luminous IR galaxies,
decreased by a similar factor in this redshift range, i.e. much more
than expected by studies at other wavelengths affected by dust
extinction. This indicates that a large fraction of present day stars
were formed during a dusty starburst event.

We estimate a contribution of ISOCAM galaxies to the peak of the CIRB
at $\lambda_{max}\simeq$ 140\,$\mu$m of (16$\pm$5) nW m$^{-2}$
sr$^{-1}$ as compared to the measured value of (25$\pm$7) nW m$^{-2}$
sr$^{-1}$ from COBE. This study therefore suggests that the ISOCAM
galaxies are responsible for the bulk of the CIRB.

We have started a systematic spectroscopic follow-up of these galaxies
with the aim of studying their physical properties and the origin of
their large SFR. Franceschini et al. (2001) estimated their baryonic
masses to be of the order of $\left< M \right> \sim 10^{11}~M_{\sun}$
by fitting their optical and near-IR luminosities with template SEDs
(from Silva et al. 1998) and assuming a Salpeter initial mass function
(from 0.15 to 100\,$M_{\sun}$). Their colors are similar to field
galaxies of similar magnitudes (Cohen 2001), hence they could not have
been selected on the basis of their optical colors. The technique
which consists in using the spectral slope in the UV domain to correct
luminosities from extinction (Meurer, Heckman \& Calzetti 1999) fails
to differentiate the luminous dusty galaxies detected with ISOCAM from
other field galaxies in the HDFN+FF (Cohen 2001). This was to be
expected since this technique only works for galaxies with $L_{\rm
IR}\lesssim~4\times10^{11}~L_{\sun}$ (Meurer et al. 2000) while most
ISOCAM galaxies are more luminous than this threshold. A property of
the ISOCAM galaxies that may give a hint on their origin is their
association with small groups of galaxies. A preliminary study of 22
ISOCAM-HDFN galaxies by Cohen et al. (2000) found that nearly all
ISOCAM galaxies belonged to small groups, while the fraction of field
galaxies with similar optical magnitudes belonging to such groups was
68\,$\%$. The study of the full sample of ISOCAM-HDFN galaxies by
Aussel et al. (2001) shows that all of them belong to physical groups,
hence suggests that dynamical effects such as merging and tidal
interactions are responsible for their large SFR.

The Space IR Telescope Facility (SIRTF) will soon provide a powerful
insight on the FIR emission of distant dusty starbursts and its
24\,$\mu$m band, less affected by confusion, should prolongate their
detection up to $z\sim2.5$. Extending the redshift range surveyed by
ISOCAM to such redshifts is particularly important to measure the
direct IR emission of the distant population of Lyman break galaxies,
whose dust extinction remains highly uncertain (Steidel et
al. 1999). However, the direct measurement of the MIR and FIR emission
of distant galaxies will only become possible with the launch of the
Next Generation Space Telescope (NGST) and Herschel (FIRST) satellite
scheduled for 2009 and 2007.

\begin{acknowledgements}
DE wishes to thank the American Astronomical Society for its support
through the Chretien International Research Grant and Joel Primack and
David Koo for supporting his research through NASA grants NAG5-8218
and NAG5-3507. We wish to thank H.Flores and A.Blain for fruitful
comments and A.Boselli for sharing his ISOCAM catalog with us. This
research has made use of the NASA/IPAC Extragalactic Database (NED)
which is operated by the Jet Propulsion Laboratory, California
Institute of Technology, under contract with the National Aeronautics
and Space Administration.
\end{acknowledgements}

\end{document}